\documentclass[aps,floatfix,superscriptaddress,reprint,showpacs,10pt,preprintnumbers,longbibliography]{revtex4-1}
\usepackage[utf8]{inputenc}
\usepackage[pdftex]{graphicx}
\usepackage{subfigure}
\usepackage{float}
\usepackage{amssymb}
\usepackage{amsmath, amsfonts}  
\usepackage{dsfont}
\usepackage{array}
\usepackage{bm}
\usepackage{mathrsfs}
\usepackage{pifont}
\usepackage{multirow}
\usepackage{upgreek}
\usepackage{diagbox}
\usepackage[normalem]{ulem}
\usepackage[dvipsnames]{xcolor}
\usepackage{soul}
\usepackage[pdftex,
            pdftitle={Warm Start of Variational Quantum Algorithms for Quadratic Unconstrained Binary Optimization Problems},
            pdfauthor={Yahui Chai, Karl Jansen, Tobias Stollenwerk},
            bookmarks,
            colorlinks,
            linkcolor=myblue,
            citecolor=mymagenta,
            menucolor=black,
            urlcolor=myblue,
            plainpages=false,
            pdfpagelabels,
            hypertexnames=false]{hyperref}
\usepackage{verbatim}
\usepackage{slashed}
\usepackage[braket,qm]{qcircuit}
\usepackage{cleveref}
\usepackage{bm}
\definecolor{mymagenta}{RGB}{200, 0, 100}
\definecolor{myblue}{RGB}{45, 48, 146}

\graphicspath{{figures/}}

\begin{document}
\title{Warm Start of Variational Quantum Algorithms for Quadratic Unconstrained Binary Optimization Problems}
\author{Yahui Chai}
\affiliation{Deutsches Elektronen-Synchrotron DESY, Platanenallee 6, 15738 Zeuthen, Germany}

\author{Karl Jansen}
\affiliation{Deutsches Elektronen-Synchrotron DESY, Platanenallee 6, 15738 Zeuthen, Germany}
\affiliation{Computation-Based Science and Technology Research Center, The Cyprus Institute, 20 Kavafi Street, 2121 Nicosia, Cyprus}

\author{Stefan Kühn}
\affiliation{Deutsches Elektronen-Synchrotron DESY, Platanenallee 6, 15738 Zeuthen, Germany}

\author{Tim Schw\"agerl}
\affiliation{Institut f\"ur Physik, Humboldt-Universit\"at zu Berlin, Newtonstr. 15, 12489 Berlin, Germany}
\affiliation{Deutsches Elektronen-Synchrotron DESY, Platanenallee 6, 15738 Zeuthen, Germany}

\author{Tobias Stollenwerk}
\affiliation{Institute for Quantum Computing Analytics (PGI-12) J\"ulich Research Center Wilhelm-Johnen-Str., 52428 J\"ulich, Germany}

\date{\today}

\begin{abstract}
    Variational Quantum Eigensolver (VQE) is widely used in near-term hardware. However, their performances remain limited by the poor trainability and are dependent on random parameter initialization. In this work, we propose a warm start method inspired by imaginary time evolution, allowing for determining initial parameters that prioritize lower energy states in a resource-efficient way. Using classical simulations, we demonstrate that this warm start method significantly improves the success rate and reduces the number of iterations required for the convergence of VQE. The numerical results also indicate that the warm start approach effectively mitigates statistical errors arising from a finite number of measurements, and to a certain extent alleviates the effect of barren plateaus.
\end{abstract}

\maketitle
\section{Introduction}

The Variational Quantum Eigensolver (VQE) is a classical–quantum hybrid algorithm that uses parametrized quantum circuits as ansätze~\cite{peruzzo2014variational, cerezo2021variational}. Because these circuits can be flexible and relatively shallow, VQE is well suited for noisy intermediate-scale quantum (NISQ) devices. The method has been applied to a wide range of problems, including eigenvalue calculations in physics and chemistry~\cite{dimeglio2023quantum, Tilly_2022} and various combinatorial optimization tasks~\cite{FGA_PRApllied, FGA_ionq, schwaegerl2023particle, Amaro_2022, PhysRevE99013304}. Despite this broad applicability, two major challenges limit its practical usefulness: statistical errors arising from finite measurement shots~\cite{Scriva_2024} and the presence of barren plateaus in the optimization landscape~\cite{McClean_2018}. Both issues can prevent the optimization from converging to the global minimum. Consequently, designing suitable ansätze and choosing effective initial parameters are important for improving VQE performance.

In this work, we investigate the quadratic unconstrained binary optimization (QUBO) problem~\cite{Kochenberger2014} and develop a warm start approach. By approximating the action of imaginary time evolution step by step, this warm start is able to provide an initialization prioritizing lower energy states and offering a more effective starting point than the typical universal superposition or random states. To effectively obtain the warm start parameters, we introduce a \emph{structure-inspired ansatz}(SIA). While the SIA alone does not outperform the hardware-efficient ansatz (HEA)~\cite{Kandala_2017} with finite shots, it gains significant improvement combined with the warm start, highlighting the importance of the initialization for the VQE. Our numerical simulation results indicate that VQE with the warm start is able to result in a higher success rate in finding the optimal solution with fewer iterations. Besides, the warm start shows a potential to mitigate the problems of statistical error and the barren plateaus.

The paper is organized as follows. Section~\ref{sec: qubo and cvar-vqe} outlines the QUBO problem and the simulation setup. Subsequently, we introduce a warm start approach in Sec.~\ref{sec: warm-start-general}, as well as low-cost simplification thereof in Sec.\ref{sec: warm-start-analytical}, which allows for determining the initial set of parameters classically. The improvement in performance of these methods is demonstrated using numerical simulation. Section~\ref{sec: discussion} shows the potential of the warm start to mitigate statistical errors and to alleviate the effect of barren plateaus. Finally, Sec.~\ref{sec: conclusion} concludes with a summary and prospects for future research.

\section{QUBO and CVaR-VQE\label{sec: qubo and cvar-vqe}}

In this paper we focus on the CVaR-VQE, a variant of the VQE proposed for combinatorial optimization problems~\cite{Barkoutsos_2020}, and apply it to QUBO problems, both of which we review below. The QUBO problem is defined by a vector of  $N$ binary variables $\bold{x} = \{x_0, x_i, \cdots x_{N-1}\}$ and the symmetric coefficient matrix $Q \in \mathbb{R}^{N \times N}$ as well as a cost function~\cite{Kochenberger2014}
\begin{equation}
    f(\bold{x}) = \bold{x}^T Q \bold{x} = \sum_{i = 0}^{N-1} \sum_{j=0}^{N-1} x_i \cdot Q_{ij} \cdot x_j.
\end{equation}
The objective is to find the assignment of binary variables which minimizes the cost function for a given $Q$. A wide range of combinatorial optimization problems can be expressed in this mathematical framework, among them the optimal flight to gate assignment at an airport, particle tracking, job scheduling, and many others~\cite{FGA_PRApllied, FGA_ionq, schwaegerl2023particle, Amaro_2022}. Every QUBO problem can be associated with a graph by interpreting the matrix $Q$ as the adjacency matrix of a weighted, undirected graph $G(V, E)$. $V$ corresponds the set of vertices of size $|V| = N$, and $E$ represents the set of edges between vertex pairs $(i,j)$ for which $Q_{ij}=Q_{ji}\neq 0$.

QUBO problems can be straightforwardly represented as Hamiltonians on $N$ qubits which are polynomials of Pauli-$Z$ operators~\cite{hadfield2021representation}. 
In particular, the corresponding Hamiltonian for $f(\bold{x})$ can be obtained by substituting the binary variables $x_i$ with the operator $\left(\mathds{1}-\sigma^z_i\right)/2$, where $\mathds{1}$ denotes the identity operator and $\sigma^z_i$ represents the Pauli-$Z$ operator acting on the $i$-th qubit.
Using this substitution one finds for the Hamiltonian up to a constant
\begin{equation}\label{eq: Hamiltonian}
    H = \sum_{i \in V} h_i \cdot \sigma^z_i + \sum_{(i,j)\in E} J_{ij} \cdot \sigma^z_{i} \sigma^z_{j},
\end{equation}
where the coefficients $h_i, J_{ij}$ are related to the original QUBO problem's $Q_{ij}$. Note that the above Hamiltonian only contains Pauli-$Z$ operators, which renders it diagonal in the computational basis, and its eigenstates are the computational basis states. This is a general property of boolean functions, as they can all be represented as diagonal matrices in the computational basis~\cite{hadfield2021representation}. For the purpose of this study, we consider random QUBOs and sample the coefficients $h_i, J_{ij}$ uniformly in the range $[-1, 1]$ with four-digit precision. 

For combinatorial optimization problems, the conditional value at risk (CVaR)~\cite{ACERBI20021487} has been proposed as a cost function for the VQE, and it was been shown to significantly enhance VQE's performance~\cite{Barkoutsos_2020, FGA_PRApllied, role_entangle}. Specifically, taking $S$ measurements in the computational basis and arranging the corresponding eigenvalues associated with the basis states sampled in ascending order, $\{E_1 \leq E_2 \leq \cdots \leq E_S\}$, the CVaR corresponds to the average of the fraction $\alpha$ of the lowest energies
\begin{equation}
    \text{CVaR}_{\alpha} = \frac{1}{\lceil \alpha S \rceil} \sum_{k = 1}^{\lceil \alpha S \rceil} E_k.
    \label{eq:CVaR}
\end{equation}
Note that in the limit $\alpha = 1$, the $ \text{CVaR}_{\alpha}$ is nothing but the usual sample mean for estimating the energy with $S$ measurements used in regular VQE. In the opposite limit, $\alpha \to 0$,  $\text{CVaR}_{\alpha}$ does nothing but selecting the measurement which resulted in the lowest energy. For $0<\alpha < 1$, the $\text{CVaR}_{\alpha}$  cost function can be understood as trying to  push more weight in the low-energy tail of the measurements, thereby enhancing computational basis states corresponding low-energy solutions in the wave function. 

In this work we focus on $\text{CVaR}_{0.01}$ unless stated otherwise. This choice balances two considerations: emphasizing the low-energy sector while maintaining sufficient sampling statistics for a stable estimate. Since our main simulations use $S = 10000$ measurement shots per iteration, setting $\alpha = 0.01$ ensures that the lowest $1\%$ of outcomes still contain 100 samples, which is enough to yield a reliable average. Such a small value of $\alpha$ therefore emphasizes the relevant low-energy sector while maintaining robustness against statistical fluctuations.

For the rest of the paper, we use the VQE with $\text{CVaR}_{\alpha}$ as a cost function as a test bed for the ansatz and the warm start procedures we develop in the following sections. To have a consistent set of test cases, we create a 100 random QUBO instances corresponding to complete graphs for each problem size $N$, where $N$ takes even numbers ranging from 12 to 24, that we use for all our numerical experiments. Throughout a VQE run, we monitor the fidelity of the final wave function with the exact solution, i.e., the probability of the optimal solutions in the quantum state, with the optimal solution obtained by brute force.  A problem solution for a particular instance is deemed as successful if the maximum fidelity throughout all iterations is larger than $1\%$. As a metric to judge the performance of the VQE, we use the success rate among the problem ensemble, that is, the fraction of instances that were successful.

\section{Warm  start inspired by Imaginary time evolution}

In this section, we first explain how the decomposition of the Imaginary time evolution (ITE) operator motivates the structure of the ansatz in Sec.~\ref{sec: SIA}. The actual warm start procedure to determine good initial parameters for this ansatz will be introduced separately in Sec.~\ref{sec: warm-start-general} and Sec.~\ref{sec: warm-start-analytical}.

\subsection{Structure-inspired ansatz}\label{sec: SIA}
Given a Hamiltonian $H$, a low-energy state can be obtained by evolving a given initial state $\ket{\psi_0}$ in imaginary time. In particular, the state tends to be the ground state, $\ket{E_0}$, in the limit
\begin{align}
    \ket{E_0} = \lim_{\tau\to\infty}\frac{\exp(-\tau H)\ket{\psi_0}}{\sqrt{\melem{\psi_0}{\exp(-2\tau H)}{\psi_0}}},
\end{align}
provided the initial state $\ket{\psi_0}$ has nonvanishing overlap with  $\ket{E_0}$. Note that the ITE operator, $\exp(-\tau H)$, is nonunitary and does not preserve the norm, thus we explicitly include the norm factor in the expression above. Although ITE provides theoretical guarantees of convergence to the global minimum~\cite{hartung2025}, it cannot be implemented directly on quantum hardware because of its nonunitary nature, motivating the development of several indirect realizations. Variational approaches approximate the ITE dynamics using parametrized circuits~\cite{Motta_2019, McArdle_2019, PRXQuantum.2.010317, Yuan2019}, which typically require nontrivial resources for accurate parameter estimation. In addition, classical–quantum hybrid Monte Carlo schemes have been proposed, where imaginary-time observables are reconstructed through importance sampling of real-time evolution~\cite{Huo2023errorresilientmonte, Han:2024qpi}. These methods avoid variational optimization and rely on repeated short-time quantum circuit executions.

For combinatorial optimization problems, where the eigenstates are given by computational basis states, the target is to find the single optimal bitstring, so it is not necessary to execute the ITE exactly to get the same resulting state. Instead, it suffices to increase the probability of the optimal (or near-optimal) bitstrings, and the distribution of higher-energy states is largely irrelevant in this case. Thus, we propose to approximate the ITE in a resource-efficient way, provide a good initial state that can guide VQE's optimization. The CVaR-VQE applied afterward will focus on refining the low-energy subspace. In this subsection, we introduce the ansatz inspired by the structure of the ITE operator, which is essential for deriving the warm start parameters that mimic the ITE process. The detailed procedure for determining these parameters will be presented in the following subsections.

For the ITE process, a suitable choice of initial state is the uniform superposition of all computational basis states, $\ket{\psi_0} = \ket{+}^{\otimes N}$,  where  $\ket{+} = \left(\ket{0} + \ket{1}\right)/\sqrt{2}$. Moreover, the terms of the Hamiltonian in Eq.~\eqref{eq: Hamiltonian} commute with each other, thus the corresponding imaginary time evolution can be decomposed exactly as
\begin{equation}
    e^{-\tau H} = \prod_{(i,j)\in E} e^{-\tau J_{ij}  \sigma_i^z \sigma_j^z} \times \prod_{i\in V} e^{-\tau  h_i  \sigma_i^z}.
    \label{eq:ITE_QUBO}
\end{equation}
Utilizing the above decomposition, we can understand the effect of the imaginary time evolution on this initial state step-by-step. Considering the second term, i.e.\ the single-body terms, we see that 
\begin{align}
    e^{-\tau  h_i  \sigma_i^z}\ket{+} = \frac{1}{\sqrt{2}}\left(e^{-\tau  h_i}\ket{0} + e^{\tau  h_i}\ket{1}\right).
\end{align}
Up to normalization, the resulting state on the right hand side can also be generated by applying the unitary rotation gate $R_y(\theta_i^{\ast})=\exp(-i\theta_i^{\ast} \sigma^y_i/2)$ to $\ket{+}$, with appropriately chosen angle $\theta_i^{\ast}$. Hence, up to normalization, we have
\begin{align}
    \ket{\psi_1} = \prod_{i\in V}R_y(\theta_i^{\ast})\ket{+}^{\otimes N} \sim \prod_{i\in V}e^{-\tau  h_i  \sigma_i^z}\ket{+}^{\otimes N} .
    \label{eq:ITE_single_qubit_terms}
\end{align}

For the two-body terms of the imaginary time evolution operator, we can proceed similarly. Starting from $\ket{\psi_1}$, the effect of the different $e^{-\tau J_{ij}  \sigma_i^z \sigma_j^z}$ in Eq.~\eqref{eq:ITE_QUBO} can again be described by a unitary $U_{ij}$ up to normalization
\begin{align}
    \ket{\psi_{k+1}} = U_{ij}\ket{\psi_k} \sim e^{-\tau \cdot J_{ij} \cdot \sigma_i^z \sigma_j^z}\ket{\psi_k} ,\quad k\geq1.
    \label{eq:ITE_two_qubit_terms}
\end{align}

The operator $U_{ij}$ might be very non-local to get the same resulting state as ITE. For simplicity, we can only consider the operators acts on the same qubits to approximate the effect of ITE. Since the ITE operator $e^{-\tau J_{ij}  \sigma_i^z \sigma_j^z}$ is a purely real matrix without complex phase, we can consider the following two-qubit operator to approximate it:
\begin{equation}\label{eq: SIA-YZ}
    U_{ij}(\boldsymbol{\theta}_{ij}) = e^{-i (\theta_{ij, 1} \cdot \sigma_i^z \sigma_j^y + \theta_{ij, 0} \cdot \sigma_i^y \sigma_j^z)/2},
\end{equation}
Although there are some other real unitaries like $ e^{-i\theta \cdot \sigma_i^x \sigma_j^y} $, $e^{-i\theta \cdot \sigma_i^x \sigma_j^y}$ can be considered, which does not have an effect in this case as shown in Appendix~\ref{app: detail-warm-start}.

Assuming that the optimal parameters $\theta_{ij,0}^{\ast}, \theta_{ij,1}^{\ast}$ to approximate the action of ITE on the current state $\ket{\psi_k}$ are available, we apply the operator $U_{ij}(\boldsymbol{\theta}^{\ast}_{ij})$ to mimic the effect of ITE on the qubit pair $(i, j)$. We then proceed to the next relevant qubit pair and repeat the same procedure. By iterating this process over all qubit pairs with non-zero couplings $J_{ij}$ in the Hamiltonian, we obtain the final state:
\begin{equation}\label{eq: warm_state}
    \ket{\psi_{(V, E)}(\boldsymbol{\theta}^{\ast})} :=
    \prod_{(i,j)\in E} U_{ij} (\boldsymbol{\theta}^{\ast}_{ij}) \times \prod_{i \in V} R_y(\theta_i^{\ast}) \ket{+}^{\otimes N},
\end{equation}
which aims to mimic the full ITE process and yields a higher probability for low-energy solutions compared to the uniform superposition. Since the resulting ansatz reflects the structure of the problem graph, we refer to it as the \emph{structure-inspired ansatz} (SIA). We emphasize that the novelty lies not in the ansatz structure itself, but in how the optimal parameters $\boldsymbol{\theta}^\ast$ are obtained to enable a warm start, this procedure is detailed in the next subsection.

\subsection{Warm start by measurement}\label{sec: warm-start-general}
To approximate ITE step by step, we determine parameters that mimic each local ITE factor. For single qubit term, the optimal parameter can be got analytically: $\theta_i^{\ast} =2\arctan\left(-\exp(-2\tau h_i)\right) + \pi/2$. In contrast, the two-body operator is more complicated and depends on the state $\ket{\psi_k}$ obtained from the previous step. Specifically, the optimal parameters of $U_{ij}(\boldsymbol{\theta}_{ij})$ in Eq.~\eqref{eq:ITE_two_qubit_terms} can be determined by maximizing the overlap between the exact ITE-evolved state and the variationally transformed state
\begin{equation}\label{eq: f-local-op}
    \begin{aligned}
        &f_{\tau, k}(\theta_{ij, 0}, \theta_{ij, 1}) \\
        &=  \bra{\psi_{k-1}} e^{-\tau \cdot J_{ij} \cdot \sigma_i^z \sigma_j^z} \cdot e^{-i (\theta_{ij,1} \cdot \sigma_i^z \sigma_j^y + \theta_{ij,0} \cdot \sigma_i^y \sigma_j^z)/2} \ket{\psi_{k-1}},
    \end{aligned}
\end{equation}
The above formula corresponds to the expectation value of a two-qubit operator, which can be expanded as a sum of Pauli strings $\hat{P} \in \{I, \sigma^x, \sigma^y, \sigma^z\}^{\otimes 2}$, and the expectation $\bra{\psi_{k-1}} \hat{P} \ket{\psi_{k-1}}$ of the individual Paul strings can be measured efficiently. As detailed in Appendix~\ref{app: detail-warm-start}, after obtaining the expectation value of those Pauli operators, the overlap $f_{\tau, k}$ for a fixed time $\tau$ only depends on the parameters $\theta_{ij,0}, \theta_{ij,1}$. The optimal parameters to maximize this overlap function can be easily found classically by solving
\begin{equation}
    \begin{aligned}        
        \theta_{ij,0}^{\ast}, \theta_{ij,1}^{\ast} &= \text{arg\ max} \ f_{\tau, k}(\theta_{ij,0}, \theta_{ij,1}).
    \end{aligned}
\end{equation}
We then apply the unitary gate $U_{ij}$ with these optimal parameters, and the state is updated by $\ket{\psi_{k+1}} = U_{ij}(\theta_{ij,0}^{\ast}, \theta_{ij,1}^{\ast}) \ket{\psi_{k}}$. This procedure is repeated for every two-body term until all corresponding gates in Eq.~\eqref{eq: warm_state} have been assigned their optimal parameters. For the rest of the paper, we dub the method discussed above \emph{warm start by measurement} (WS-M) to distinguish it from the approximate warm start technique proposed in the next section. 

Note that this warm start will only provide an approximation to the ITE. Since we have chosen a specific ansatz structure for the two-qubit gates, it can not produce the effect of the ITE operator up to arbitrarily large $\tau$. For intermediate values of $\tau$, we expect that the WS-M approach can provide a good initial state for the VQE. 

\begin{figure}[!htp]
    \centering
    \includegraphics[width = 0.48\textwidth]{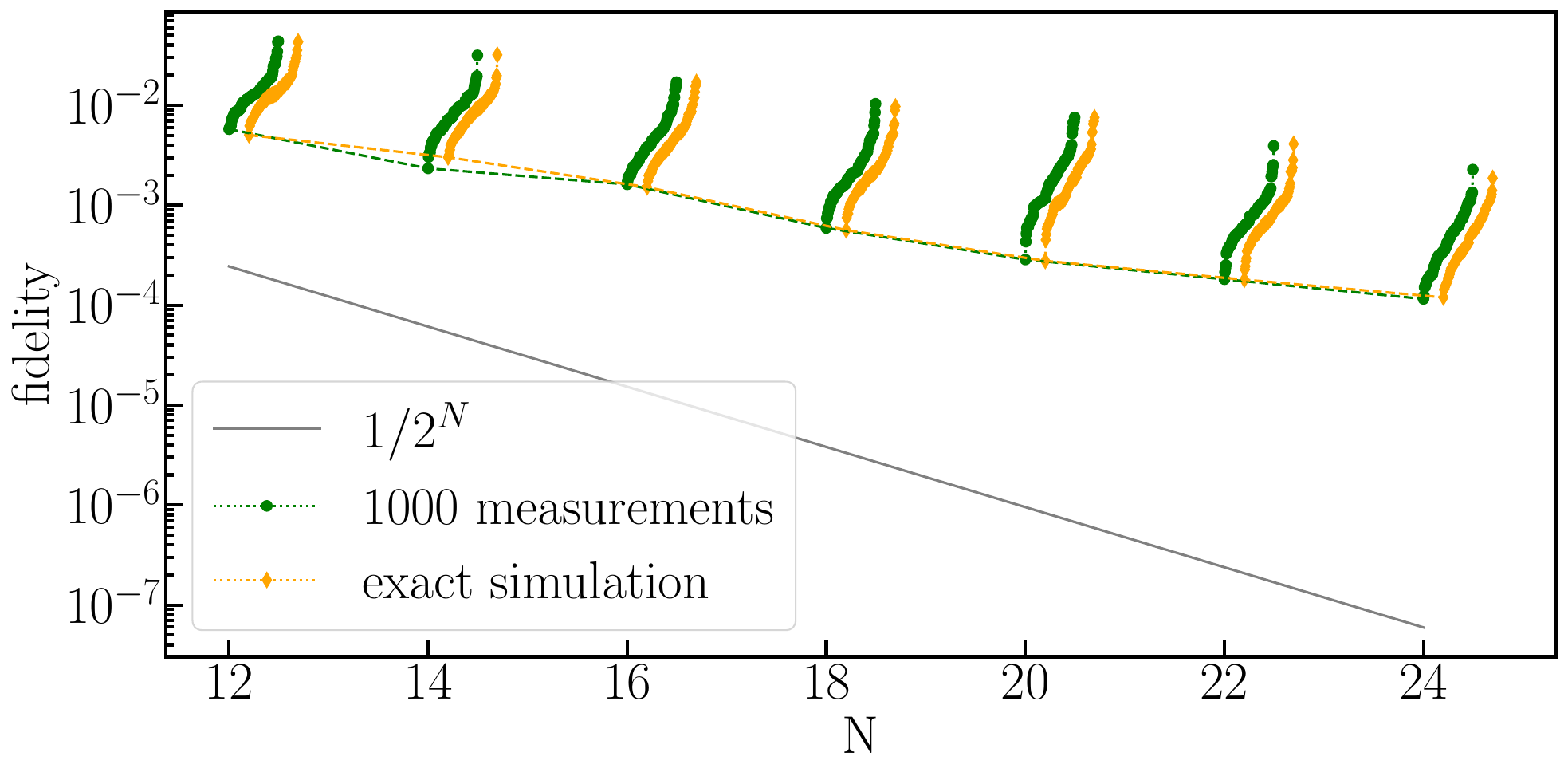}
    \caption{Fidelity of the warm start state that mimics ITE with $\tau = 0.2$. The yellow diamonds represent the initial state following the WS-M technique using a state vector simulation, corresponding to an infinite number of measurements when estimating the expectation values. The green dots correspond to the data obtained using 1000 measurements for estimating each of the local Pauli operators. For each system size, the fidelity of 100 random QUBO problems are presented, sorted in ascending order and slightly offset on the $x$-axis for viewer convenience. The dashed line connects the minimum fidelity at each system size to guide the eye. The gray line shows the fidelity between the solution and the uniform superposition.}
    \label{fig: warm-start-fidelity}
\end{figure}

Figure~\ref{fig: warm-start-fidelity} demonstrates the efficiency of our warm start approach, by showing the fidelity of the state with initial parameters according to the WS-M procedure compared to the uniform superposition.
Across all random QUBO instances we study, the warm start consistently achieves a significantly higher fidelity than the uniform superposition. For clearer visualization, we connect the minimum fidelity at each system size with an orange dashed line in Fig.~\ref{fig: warm-start-fidelity}. Compared to the baseline given by the uniform superposition, these minimum fidelities decrease much more slowly with system size and exceed the baseline by several orders of magnitude across the entire range studied. Besides, using a modest budget of 1000 measurements to estimate the expectation value of the local Pauli operators in Eq.~\eqref{eq: f-local-op}, the simulated data taking into account a finite number of measurements is close to the exact state vector simulation. Remarkably, this consistency does not decay with the number of qubits, which indicates the necessary number of measurements does almost not increase with the number of the qubits for the entire range of problem sizes we consider. This is likely because the WS-M only requires the estimation of expectations of two-qubit operators. 

To show that the WS-M not only increases the fidelity of the initial state with the optimal solution, but generally also leads to a better VQE performance, we also compare the performance of the VQE using the SIA ansatz with and without warm start. For the case without warm start, all parameters in the circuit Eq.~\eqref{eq: warm_state} are set to zero initially, which has a comparable performance as the hardware-efficient ansatz with the same initialization as shown in Appendix.~\ref{app: benchmark_ansatz}.  The classical minimization is performed with constrained optimization by linear approximation (COBYLA)~\cite{Powell1994}, with the maximum number of iterations set to $50\times N$, to provide a benchmark with a number of cost function evaluations realistic on current quantum hardware.

\begin{figure}[htp!]
    \centering
    \includegraphics[width = 0.48\textwidth]{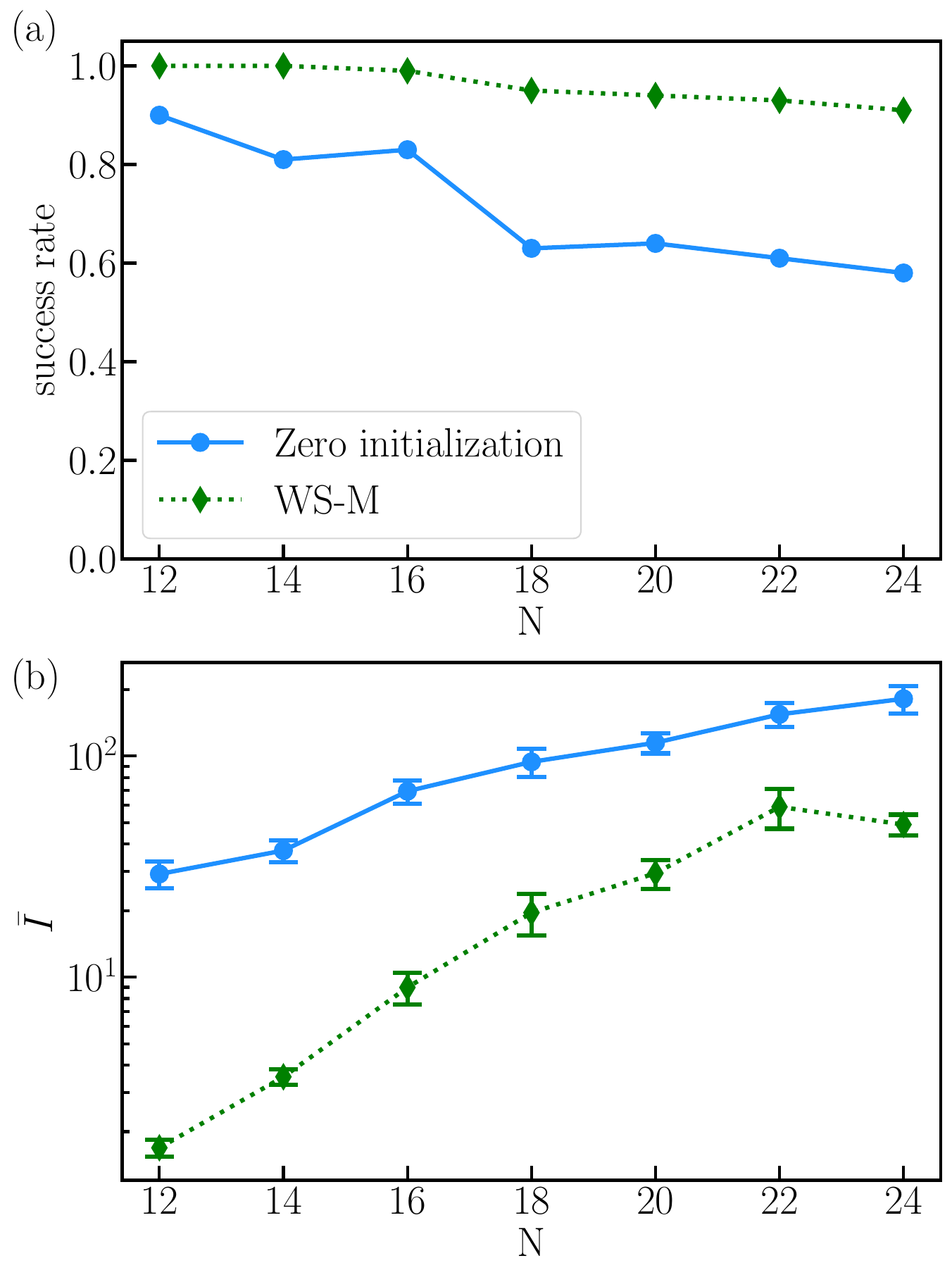}
    \caption{(a) Success rate of CVaR-VQE with $\alpha = 0.01$ and (b) the averaged iterations to achieve the fidelity threshold $1\%$, with the errorbar representing the standard error. In both panels, the green diamond represents the data using the WS-M with $\tau = 0.2$. For comparison, the blue dots show the outcomes when the VQE is initialized with a uniform superposition, i.e., all ansatz parameters are set to zero.
 The number of measurements for the warm start parameter estimation is 1000, and 10000 measurements are used for estimating the CVaR value in each VQE iteration for both cases, with or without the warm start. 
    }
    \label{fig: warm_start_sr_shots}
\end{figure}
As shown in Fig.~\ref{fig: warm_start_sr_shots}(a), warm-started VQE has a much higher success rate compared to starting VQE with the uniform superposition state. The separation in success rate between the cases grows slightly as the problem size increases. In addition, for the successful instances, we also record the minimal number of iterations to achieve the fidelity threshold $1\%$. Figure~\ref{fig: warm_start_sr_shots}(b) shows the average number of iterations, $\bar{I}$, among the successful instances to reach the desired fidelity threshold. As the figure reveals, $\bar{I}$ is noticeably smaller when using the warm start initial state, indicating a faster convergence rate of the VQE with warm start parameters towards the optimal solution. Specifically, for 24 qubits, $\bar{I}$ of VQE without warm start is around $180$, whereas with warm start, it reduces to about $50$, thus achieving the fidelity threshold up to 3.6 times more rapidly.  While our system sizes are still relatively small, these results strongly suggest that the warm start enhances VQE convergence and robustness. We plan to explore its scaling properties in follow-up work among larger system sizes. 

For completeness, we also examined whether an appropriate warm start initialization could improve the performance of the HEA ansatz, as discussed in Appendix~\ref{app:hea_ws_appendix}. When we use the solution of the relaxed QUBO problem~\cite{Egger2021warmstartingquantum} as the warm start, the initial state is slightly better than the uniform superposition, but this does not lead to improved VQE optimization outcomes for HEA. This contrast further highlights the importance of the SIA ansatz together with the physically motivated warm start strategy employed in our work.

\subsection{Warm start by approximation\label{sec: warm-start-analytical}}
\begin{figure}[!htbp]
    \centering
    \includegraphics[width = 0.48\textwidth]{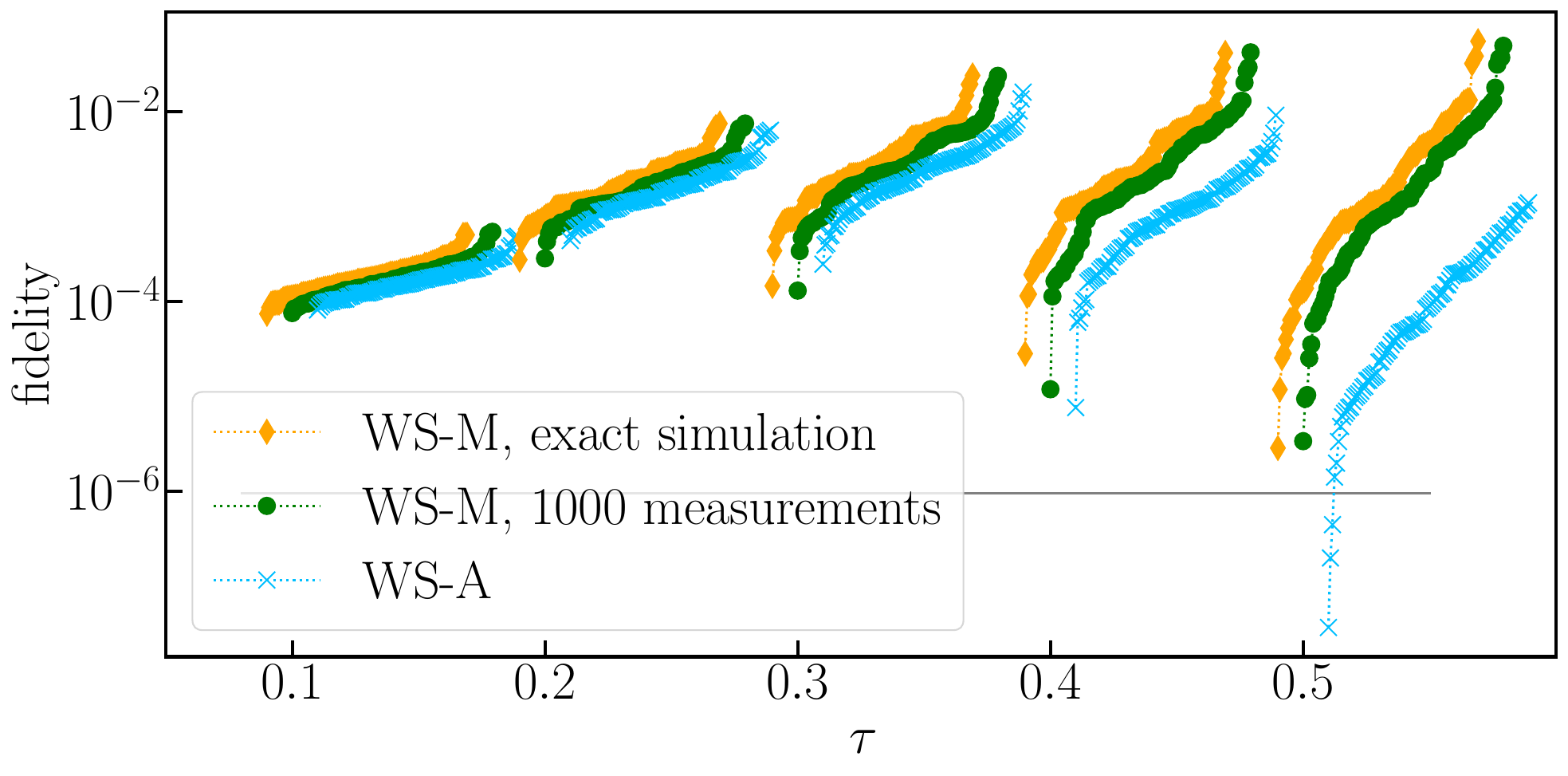}
    \caption{Comparison of warm start fidelity for 20 qubits QUBO problems. The fidelity of the WS-M method introduced in Sec.~\ref{sec: warm-start-general}, is presented by the yellow diamond for state vector simulation and green dots for simulation with 1000 measurements. The fidelity resulting from the WS-A method is denoted by blue crosses. The gray line represents the fidelity of the uniform superposition, which is $1/2^{20}$.}
    \label{fig: gi_compare_measure_analy}
\end{figure}

The WS-M approach allows for improving the performance of the VQE using $\text{CVaR}_\alpha$, as we have demonstrated in the last section. However, this comes at the expense of having to perform additional measurements. Here we demonstrate an approximate way for warm starting the VQE. As shown in Appendix~\ref{app: detail-warm-start}, the expectation value of a Pauli operator $\hat{P}$ with respect to the intermediate state $\ket{\psi_{k-1}}$ can be approximated as:
\begin{align}
    \bra{\psi_{k-1}} \hat{P} \ket{\psi_{k-1}} = \  ^{N \otimes} \bra{+} \hat{P} \ket{+}^{\otimes N} + \mathcal{O}(\tau^2).
    \label{eq:Pauli_expectation_approximation}
\end{align}
Hence, provided $\tau$ is small, this allows for approximating the warm start parameters by evaluating the expectation values of Pauli operators $\hat{P}$ in the state $\ket{+}^{\otimes N}$. These can be obtained analytically, and evaluate either to $0$ or $\pm1$, so that an approximation of the warm start parameters can be obtained without running the quantum circuit. Hence, we dub this technique \emph{warm start by approximation} (WS-A). As Fig.~\ref{fig: gi_compare_measure_analy} shows, for values of $\tau \leq 0.3$, the WS-A yields fidelities with the exact solution comparable to those of the WS-M discussed previously. 
Only for larger values of $\tau$, the fidelity of the state initialized with WS-A decreases, which can be attributed to the increasing error term $\mathcal{O}(\tau^2)$ in estimating the expectation value of Pauli operators using state $\ket{+}^{\otimes N}$. Further comparisons between these two warm start approaches, in particular, for a larger number of ansatz layers are presented in Appendix~\ref{app: layers}, revealing that the WS-M approach has a more stable performance using a larger number of layers.
    
\begin{figure}[!htbp]
    \centering
    \includegraphics[width = 0.48\textwidth]{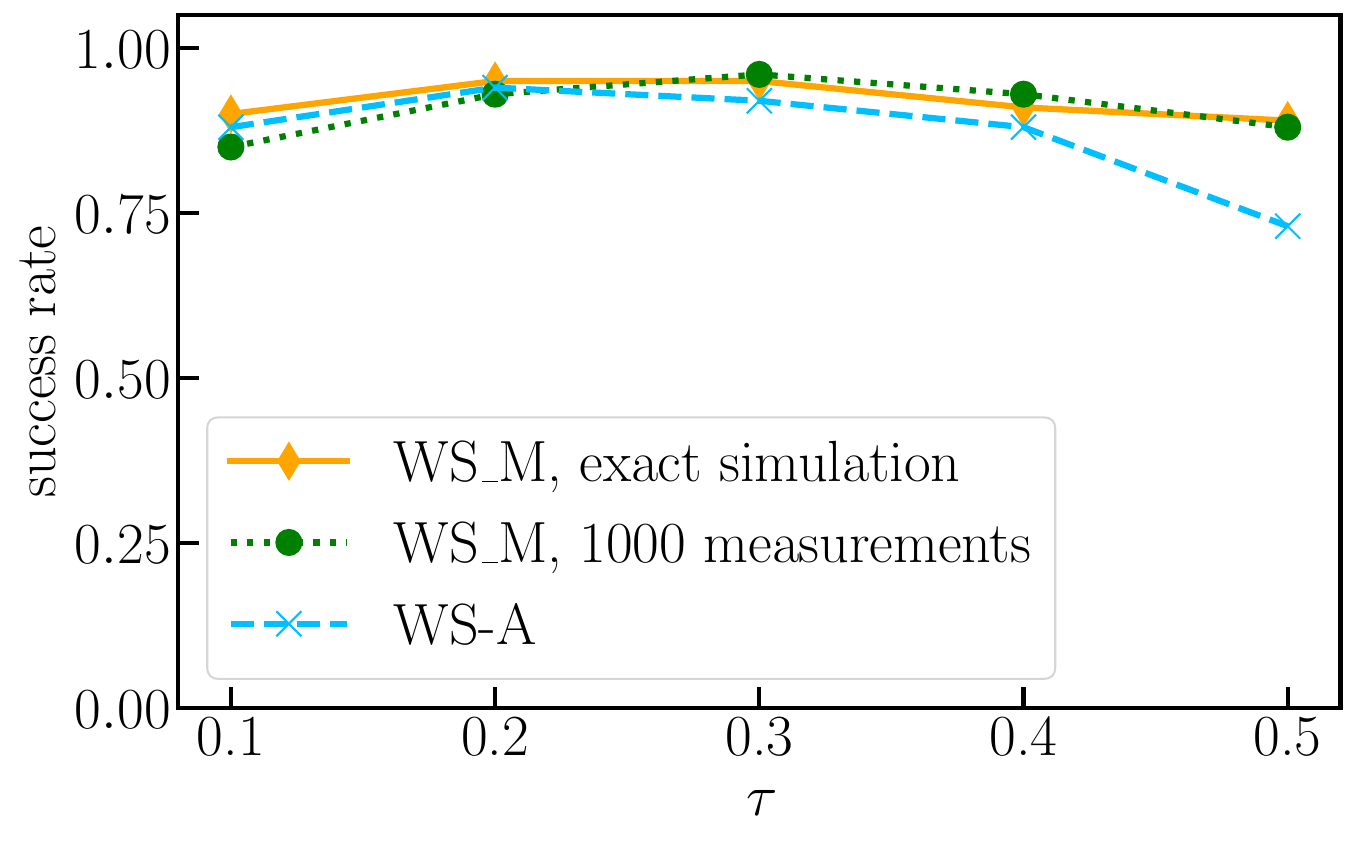}
    \caption{Success rate of CVaR-VQE with $\alpha = 0.01$ initiated with different warm start parameters for 20 qubits. Each VQE iteration involved 10,000 measurements for estimating CVaR across all three warm start approaches.}
    \label{fig:sr_gi_tech_tau}
\end{figure}

In Fig.~\ref{fig:sr_gi_tech_tau}, we show the success rate of the VQE initialized with the warm start parameters derived from different approaches for $20$ qubits. This outcome exhibits a pattern consistent with that observed in Fig.~\ref{fig: gi_compare_measure_analy}: the success rate for the WS-M approach with 1000 measurements is on par with that of infinite shots, and the approximation approach has a comparable performance when $\tau$ is small, but performs worse when $\tau$ increasing. Subsequent sections will focus on $\tau=0.2$, where both warm start methods show promising performance with a single layer. For stable performance across different layers, we will use the WS-M method to generate good initial parameters for VQE.

As our results demonstrate, for a single layer there is essentially no difference in performance between the different warm start approaches, and for the following section, we sill simply focus on the approach WS-M.

\section{Effects of statistical errors and avoidance to barren plateaus\label{sec: discussion}}

To elucidate the factors underlying the effectiveness of the warm start strategy, we first analyze the impact of statistical errors arising from a finite number of measurements. We then examine the cost concentration of the SIA ansatz and confirm that the ansatz itself exhibits barren plateaus~\cite{Arrasmith_2022} when initialized with random parameters. Importantly, the warm start procedure does not eliminate barren plateaus in general. Instead, it provides an initialization with gradients that do not decay with the system size, as shown in Sec.~\ref{subsec: discuss barren plateau}. This advantageous starting point enables the optimization to avoid flat regions of the landscape during the early stages of VQE and leads to an efficient minimization of the cost function, as illustrated in Sec.~\ref{subsec: discuss barren plateau}.

\subsection{Statistical errors due to a finite number of measurements}\label{subsec: discuss finite shots}

To study the effect of a finite number of measurements, we utilize the QUBO instances with 18 qubits as a test bed. For these instances, the statevector simulation of the VQE shows nearly perfect success rates using the SIA ansatz: $99\%$ for all initial parameters set to zero, $100\%$ for warm start. Thus, this setting serves as a testbed for analyzing how statistical errors influence success rates in each methodology.

\begin{figure}[htp!]
    \centering
    \includegraphics[width = 0.48\textwidth]{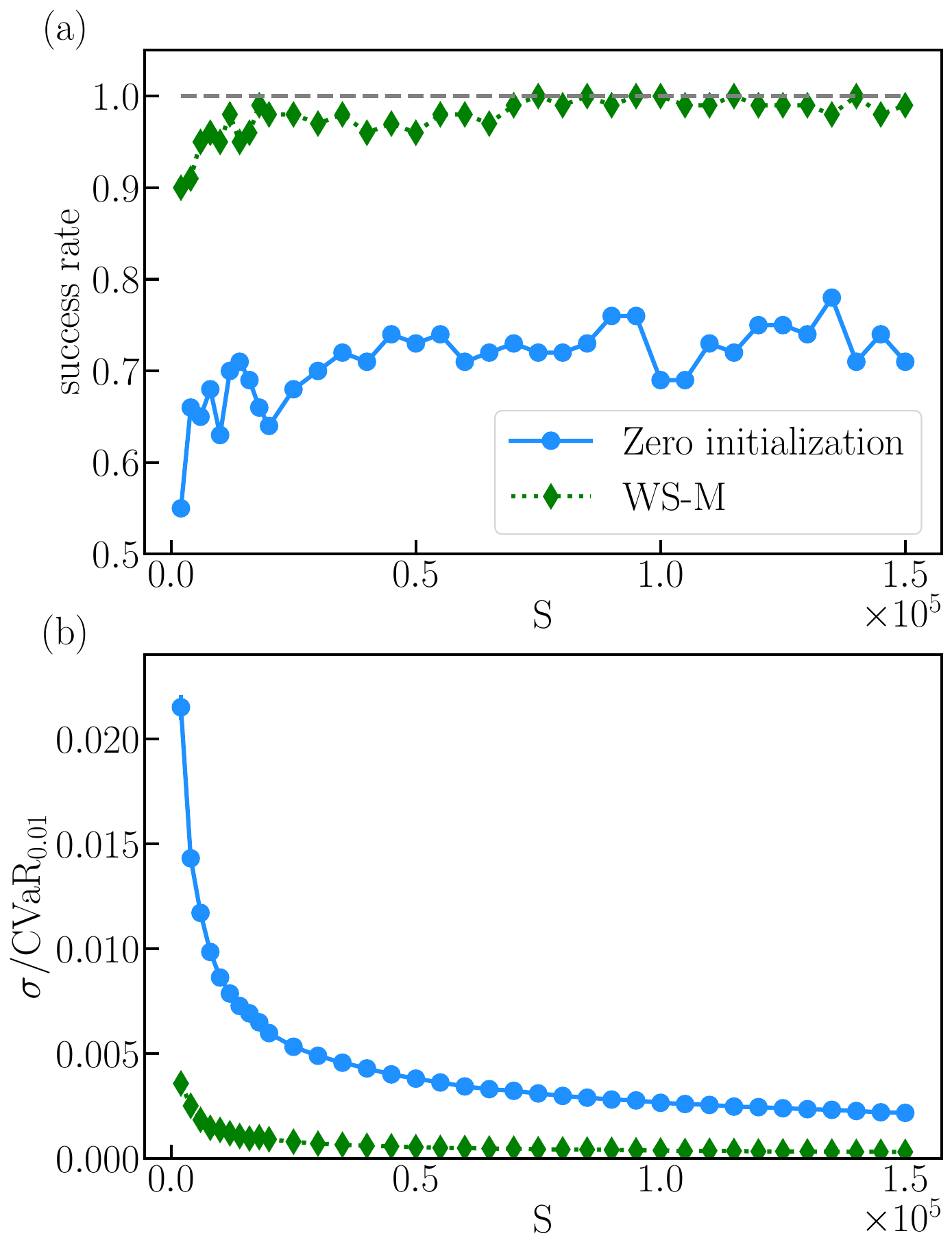}
    \caption{Scaling of the success rate with number of measurements of the CVaR-VQE with $\alpha=0.01$ (a) and relative standard error (b) for 18 qubits with averaged over 100 QUBO instances. The $x$-axis represents different numbers of measurements taken for determining the cost function in each iteration, ranging from 2000 to 150000. The blue dots represent the data for the VQE with all parameters in SIA-YZ ansatz initialized as zero, the green diamonds the results for initializing the ansatz using the warm start parameters. The grey dashed line in panel (a) indicates a success rate $100\%$.}
    \label{fig: shots_scaling}
\end{figure}

Figure~\ref{fig: shots_scaling}(a) illustrates how the success rate varies with the number of measurements used in every iteration, clearly indicating that the VQE with a warm start is more robust against the finite shots.
Using the warm start, one consistently observes success rates close to $100\%$ as soon as the number of measurements is larger than 70000. Even for as little as 2000 measurements, one can achieve a high success rate of $90\%$. In contrast, the VQE without a warm start achieves only maximum success rates of approximately $70\%$, even for 150000 measurements. To obtain further insight into the mechanism behind the robustness of the warm start, we calculate the standard error of the expectation value for the cost function $\text{CVaR}_{\alpha}$ with coefficient $\alpha$ and a number of measurements $S$. The standard error is given by
\begin{equation}
    \sigma = \sqrt{\frac{\sum_{k=1}^{\lceil \alpha \cdot S \rceil } \left(E_k - \text{CVaR}_{\alpha}(E)\right)^2}{\lceil \alpha \cdot S \rceil \left(\lceil \alpha \cdot S \rceil - 1 \right)}},
\end{equation}
where $\alpha = 0.01$ in this study and the obtanined from each of the measurements are sorted in ascending order: $E_1 \leq E_2 \leq \cdots \leq E_S$. A larger standard error indicates greater uncertainty in $\text{CVaR}_{0.01}$ when estimating the cost function with finite number of measurements. This might mislead the classical optimizer when selecting the a new set of parameters to minimize the cost function value. The relative standard error $\sigma/\text{CVaR}_{0.01}$ of the 10th iteration in VQE is plotted in Fig.~\ref{fig: shots_scaling}(b), demonstrating that the standard error is significantly reduced when utilizing a warm start. This might explain the improved performance of the warm start with finite measurements. In turn, the reduced standard error is as a result of the warm start procedure mimicking the ITE initially pushing more weights into the low-energy states, resulting in a more concentrated lowest-$\alpha$ tail of the measured eigenvalue distribution.

\subsection{Avoidance to barren plateaus\label{subsec: discuss barren plateau}}
    
The problem of barren plateaus is one of the most difficult problems for the VQE, as it can prevent the trainability of ansätze~\cite{McClean_2018}. Thus, it is pertinent to investigate if the SIA ansatz and warm start can mitigate this problem. To this end, we examine the scaling of the cost concentration with the number of qubits for the SIA ansatz, which was proven to be exponentially decaying if the cost landscape has the problem of barren plateaus~\cite{Arrasmith_2022}. 

To study to what extent the SIA ansatz is affected by barren plateaus with increasing circuit depth, we consider the ansatz with a number of layers $L\geq 1$ which is given by
\begin{equation}\label{eq: SIA-ansatz-layers}
    \prod_{l=1}^{L} \left( \prod_{(i,j)\in E} U_{ij} (\boldsymbol{\theta}_{ij}^l) \times \prod_{i \in V} R_y(\theta_i^l) \right) \ket{+}^{\otimes N}.
\end{equation}
Furthermore, in this section, besides $\alpha=0.01$, we also consider the $\alpha = 0.1$, because some instances for $N = 12, 14, 16$ might gain fidelity larger than $0.01$ in the warm start, where the global minimal is already achieved for $\alpha=0.01$, resulting in zero gradient. This could cause an increase in gradient as the number of qubits increases, so we also consider $\alpha=0.1$ to verify the scaling of the gradients with the problem size.

Firstly, we explore the variance of the cost difference with increasing number of ansatz layers. For each random QUBO instance and each number of layers $L$, we sample 2000 random parameter vectors, $\boldsymbol{\theta} = \{\theta_i^l,  \boldsymbol{\theta}_{jk}^l| i\in V, (j, k)\in E, 1 \leq l \leq L\}$, and evaluate the variance of the cost difference as in Eq.~(13) in Ref.~\cite{Arrasmith_2022}
\begin{equation}
    \text{Var}(\Delta C) = \text{Var}_{\boldsymbol{\theta}}[C(\boldsymbol{\theta}) - \mathbb{E}_{\boldsymbol{\theta}}[C(\boldsymbol{\theta})]].
\end{equation}
As throughout the rest of the paper, the cost function $C(\boldsymbol{\theta})$ is given by the CVaR$_{\alpha}$ in Eq.~\eqref{eq:CVaR}. 

\begin{figure}[htp!]
    \centering
    \includegraphics[width = 0.48\textwidth]{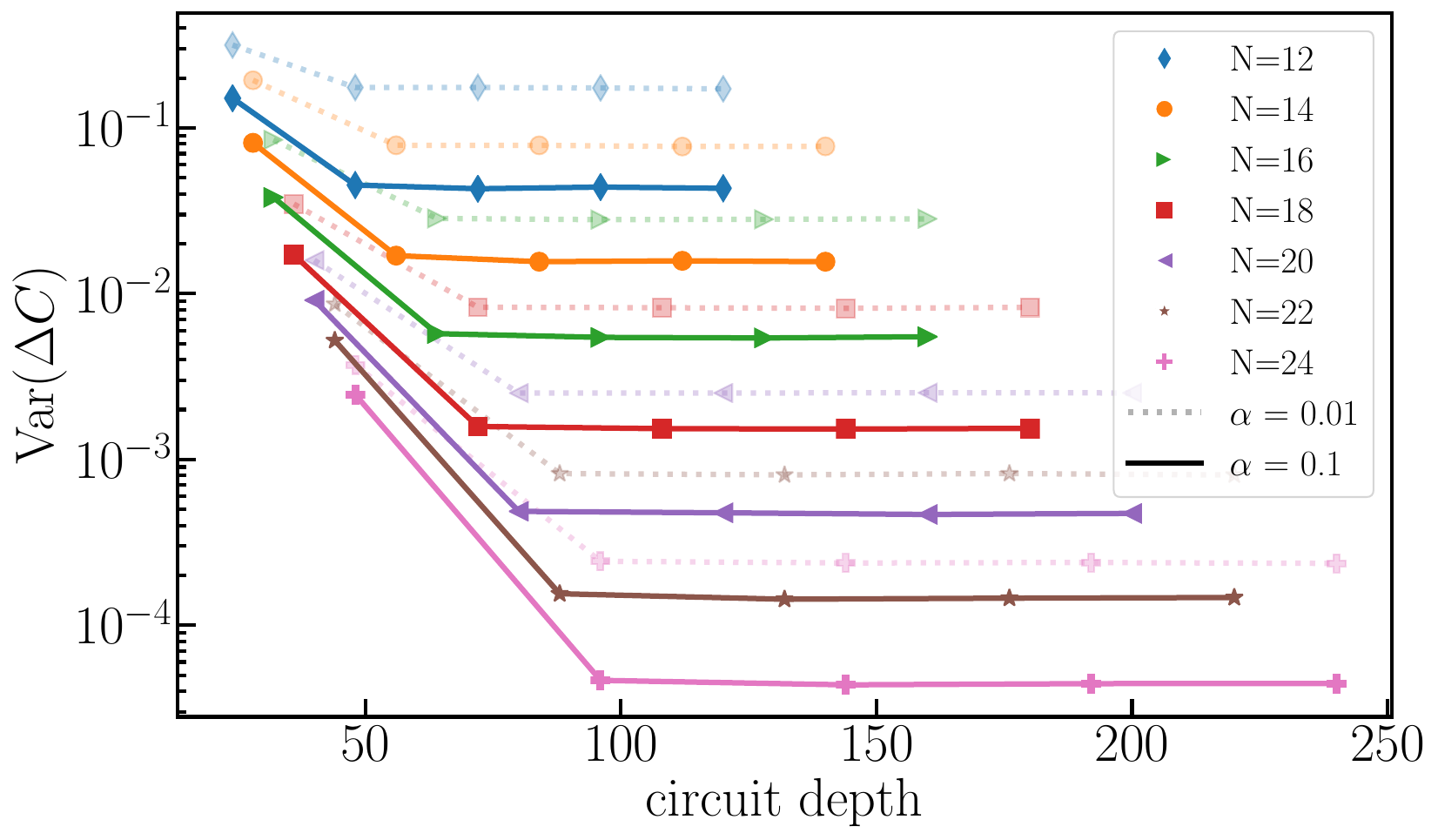}
    \caption{Cost concentration of the SIA ansatz for an even number of qubits ranging from 12 to 24. For each problem size, the value of $\text{Var}(\Delta C)$ is averaged over 10 random instances. The $x$-axis represents the circuit depth (layers of CNOT gates), which equals $2 N L$ if considering an all-to-all connected hardware, as explained in Appendix~\ref{app: ansatzes}.}
    \label{fig: bp_layers}
\end{figure}

The numerical simulation result is shown in Fig.~\ref{fig: bp_layers}. As the depth of a single layer of our ansatz circuit depends on the number of qubits $N$, we show $\text{Var}(\Delta C)$ as a function of circuit depth for better comparability of results for different problem sizes. Looking at Fig.~\ref{fig: bp_layers}, we initially observe an exponential decay, before reaching a plateau at $L \geq 2$. This indicates the SIA ansatz will form a unitary 2-design upon reaching sufficient depth~\cite{McClean_2018}. In addition, we observe that the value of the plateau reached for a fixed problem size decays exponentially with the number of qubits $N$. Our results indicate that the SIA generally shows barren plateaus as $L$ increases. 

\begin{figure}[htp!]
    \centering
    \includegraphics[width = 0.48\textwidth]{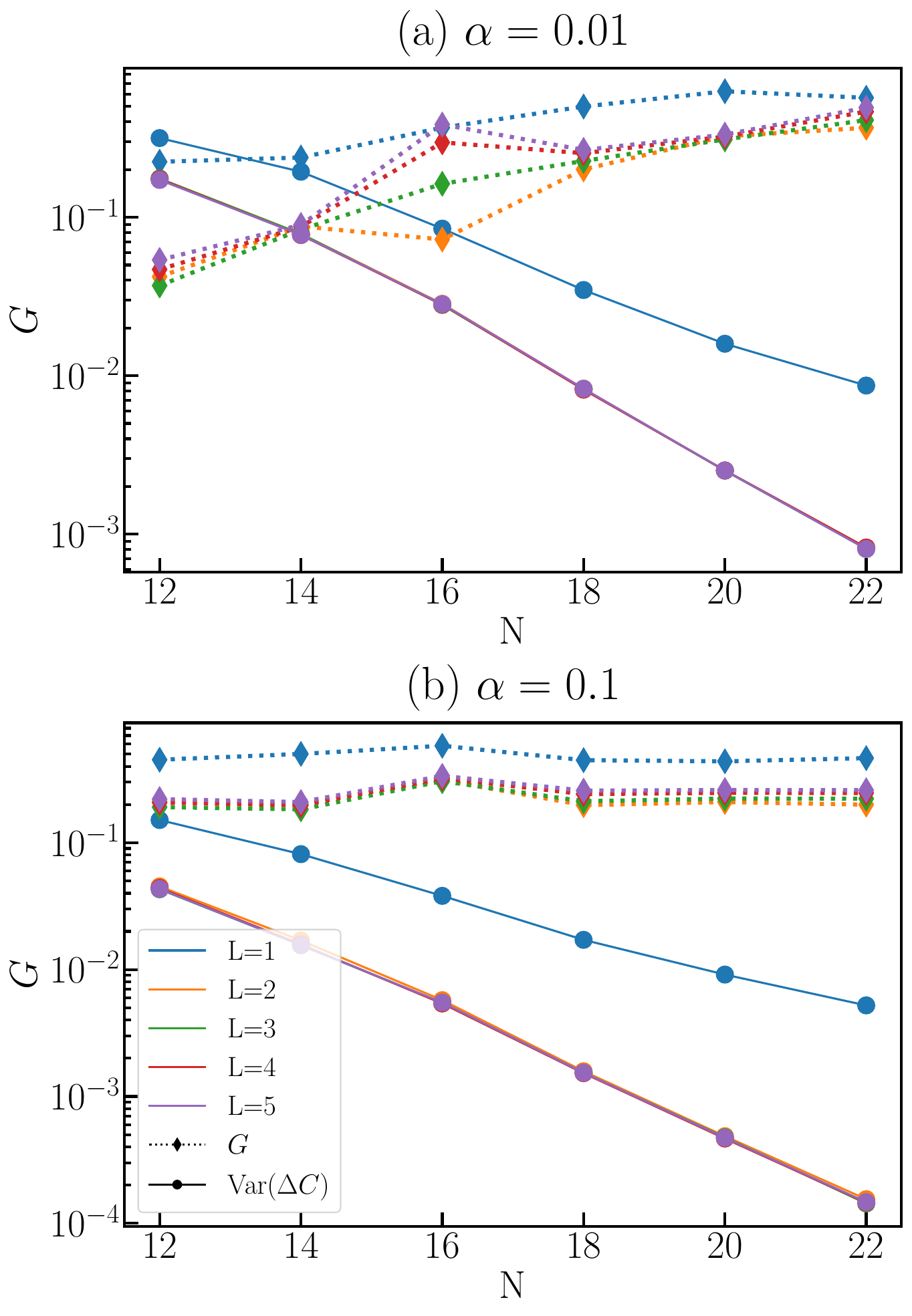}
    \caption{Scaling of the gradient obtained using the warm start parameters and cost concentration with the number of qubits, with (a) $\alpha=0.01$ and (b) $\alpha=0.1$. The dashed lines represent the square of the gradient for warm start parameters, and the solid lines represent the cost concentration. Different colors indicate different layers, and the cost concentration for $L \geq 2$ are degenerated in the plot.}
    \label{fig: bp_qubits}
\end{figure}

Furthermore, we also study the gradient observed using the warm start. Note that in the case of $l \geq 2$, the parameters of single-qubit gates in the WS-M approach should also be determined by maximizing the overlap function similar to Eq.~\eqref{eq: f-local-op}, because the state they are acting on is no longer a product state anymore. To facilitate a direct comparison with the cost concentration, which correlates with the gradients' variance, we explore the squared gradient values normalized by the dimension of the parameter space, $\text{dim} (\boldsymbol{\theta})$, for a given instance
\begin{equation}
    G = \frac{1}{\text{dim} (\boldsymbol{\theta})} \sum_{i=1}^{\text{dim}(\boldsymbol{\theta})} \left( \partial_i C(\boldsymbol{\theta}) \right)^2.
\end{equation}
The data for $G$ are shown in Fig.~\ref{fig: bp_qubits} for various problem sizes. As the figure reveals, the gradient obtained using the warm start parameters is not decaying with an increasing number of qubits. In particular, for the smaller value of $\alpha = 0.01$ we observe an increase in gradient magnitude. This can be primarily attributed to the high fidelity of the warm start for smaller system sizes. As we have shown in Fig.~\ref{fig: warm-start-fidelity}, for $N=12$ most instances have a fidelity exceeding $0.01$ using the warm start parameters. Thus, for $\alpha=0.01$ the cost function is essentially already optimal, and the gradient will be zero. As the number of qubits increases, the fidelity of the warm start eventually decreases below $\alpha$, leading to a nonvanishing gradient and, thus, to the observed increase in gradient magnitude, as can be seen in Fig.~\ref{fig: bp_qubits}(a). For the data with $\alpha=0.1$, shown in Fig.~\ref{fig: bp_qubits}(b), we observe essentially a constant value of the gradient throughout the entire range of system sizes we study. This is a result of the fact that in this case the warm start does not reach fidelities lager than $\alpha$ (see Fig.~\ref{fig: warm-start-fidelity}), and the cost function for smaller $N$ is not close to the optimal value. In both case, $\alpha=0.01$, $0.1$, increasing the number of layers in the ansatz from 1 to 2 will result in a better warm start, resulting in a smaller value of the gradient. However, increasing the number of layers further beyond 2 does not necessarily enhance fidelity and consequently does not reduce the value of gradient further, explaining why the data for $L \geq 2$ is approximately the same for large $N$. In summary, initializing the ansatz using the warm start method does not lead to exponentially vanishing gradients, while naively increasing the number of layers will not improve the performance systematically.

The warm start provides a set of initial parameters with a large gradient, ensuring a trainable first step in the VQE. However, there are still questions about the trainability of subsequent iteration steps, and whether the warm start will lead the optimization to global or local minima. The high success rate and fast convergence to the optimal solution shown in Fig.~\ref{fig: warm_start_sr_shots} indicates that the warm start parameter should benefit these problems. To demonstrate this in a more straightforward way, we choose a random instance with $N=24$ qubits and a single ansatz layer to examine the optimization process of the VQE with different parameter initialization strategies, aiming to confirm that the warm start provides a starting point that is easier to be optimized to the optimal solution. For simplicity, we only present the approximation ratio during the VQE of $\alpha=0.01$, the conclusions remain consistent for $\alpha=0.1$. 
\begin{figure}[htp!]
    \centering
    \includegraphics[width = 0.48\textwidth]{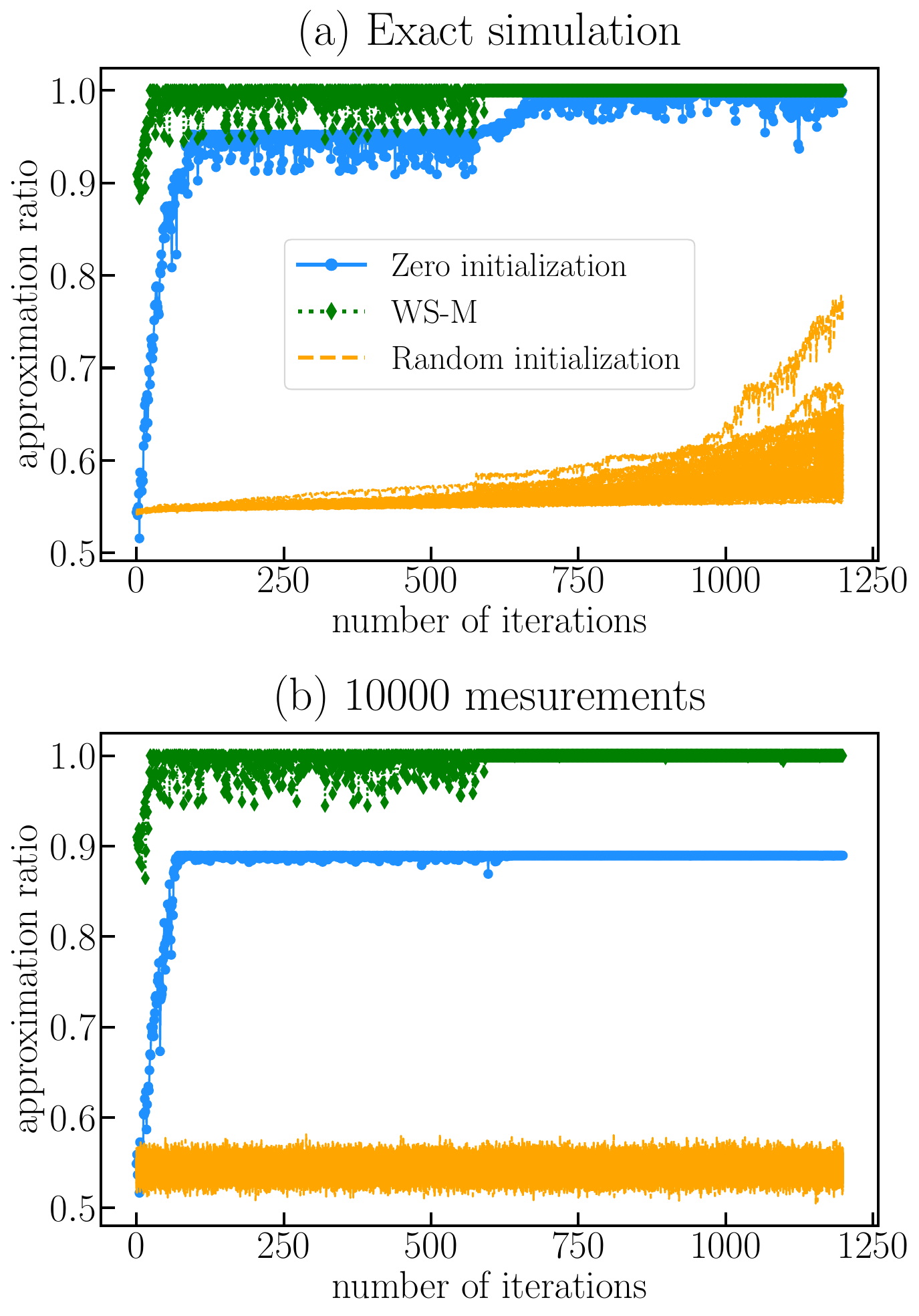}
    \caption{The approximation ratio during VQE process with different starting points of $L=1$. The $x$-axis represents the number of iterations and the $y$-axis represents the approximation ratio CVaR$_{0.01} / E_{\mathrm{opt}}$, with $E_{\mathrm{opt}}$ being the minimal cost function value obtained by the brute force search. Blue dots correspond to the VQE process starting at the uniform superposition state. The green diamonds represent that starting with the warm start. The orange dashed lines correspond to VQE processes starting with 100 random initial parameters.}
    \label{fig: bp_vqe}
\end{figure}
As shown in Fig.~\ref{fig: bp_vqe}, a random choice of initial parameters essentially leads to an untrainable ansatz, especially in the case of using a finite number of measurements. Even with an exact state vector simulation (see Fig.~\ref{fig: bp_vqe} (a)), corresponding to an infinite number of measurements, most instances show barely any improvement of the cost function during the training process. The few instances that do show some improvement converge rather slow and stay far away from the minimum cost function value over the course of 1250 iterations. As soon as we use a finite number of measurements $S=10000$ (c.f.\ Fig.~\ref{fig: bp_vqe} (b)), we observe no substantial improvement of the cost function throughout the training process for randomly chosen initial parameters. This observation aligns with our findings on barren plateaus for the SIA ansatz discussed previously. 

Initializing with all parameters being zero, corresponding to the uniform superposition as initial state, will result in a large gradient in the beginning, which enables the optimizer to quickly find a direction where the cost function decreases. However, even with an infinite number of measurements it takes a long time to get close to the global minimum and the optimization process revolves around a local minimum for a long time, as Fig.~\ref{fig: bp_vqe}(a) demonstrates. Using a finite number of measurements, Fig.~\ref{fig: bp_vqe}(b) illustrates that this initialization tends to get stuck in local minima, explaining the low success rate observed previously in Fig.~\ref{fig: success_rate}(b). This issue can be attributed to the large statistical errors as discussed in Sec.~\ref{subsec: discuss finite shots}. 

In contrast, initializing parameters using the warm start method results in an already good approximation ratio initially, and the optimization procedure reliably converges to the optimal solution, in scenarios with both infinite and finite shots, as the results in Fig.~\ref{fig: success_rate} reveals. This indicates that our warm start is able to provide a good starting point in the cost function landscape that is close to the narrow gorges in which the cost function is concentrated and can help to overcome trainability issues.

\section{Summary and outlook\label{sec: conclusion}}
In this work, we propose a resource-efficient warm-start method that mimics the imaginary time evolution of a uniform superposition over all computational basis states. Performing classical simulations with up to $24$ qubits we numerically benchmarked the performance of the warm start for the VQE using the conditional value at risk as a cost function. Our data demonstrate that the warm start procedure allows for obtaining a good performance, even with a modest number of measurements per iteration in the VQE. The boost in performance through the warm start method can be primarily attributed to the fact that the initial state generated has a dominant component of low-energy states, resulting in a reduction of statistical errors when estimating expectation values with a finite number of measurements. Moreover, we examined how the warm start parameters avoid barren plateau regions. While we observe that the ansatz in general suffers from exponential cost concentration, our numerical results indicate that the warm start provides a point in parameter space sufficiently close to one of the narrow gorges where the cost function does not look flat. 

It is worth mentioning that in our warm start approach, we determined the optimal parameter of each gate in the ansatz individually while keeping the others fixed to mimic the ITE. While this is the most resource-efficient approach in terms of overhead require for the warm start, optimizing multiple parameters at the same time might allow for faithfully emulating the effect of imaginary time evolution up to longer time-scales. This could provide an initial state that has even larger overlap with the solution as our current technique. Evaluating the trade-off between the cost of obtaining good warm start parameters and the resulting performance enhancement in the subsequent VQE will be an interesting aspect for further studies. Beyond VQE, the proposed warm-start strategy may be applicable to other quantum algorithms. For example, it could serve as an initial state for Grover search, increasing the initial overlap with the target solution and reducing the number of required iterations. Exploring such extensions is another promising avenue for further investigation.

Moreover, in our simulations considered an ideal quantum device performing a finite number of measurement. For the future it would be interesting to realize the simulation on current noisy quantum hardware and to investigate if the warm start shows a similar performance improvement in the presence of noise and in combination with state of the art error mitigation methods~\cite{Endo2018, pauli_twirling_for_coherent_noise, Funcke2020, trex, dynamical_decoupling}. In addition, it is an interesting question different kinds of hardware noise affect the warm start procedure and how these could possibly be mitigated.


\subsection*{Funding}
This work is funded by the European Union's Horizon Europe Framework Programme (HORIZON) under the ERA Chair scheme with grant agreement no.\ 101087126 and by the Deutsche Forschungsgemeinschaft (DFG, German Research Foundation) - Project-ID 429529648 - TRR 306 QuCoLiMa (``Quantum Cooperativity of Light and Matter''). This work is supported with funds from the Ministry of Science, Research and Culture of the State of Brandenburg within the Centre for Quantum Technologies and Applications (CQTA).  The funding bodies had no role in the design of the study, data collection and analysis, or writing of the manuscript.
\begin{center}
    \includegraphics[width = 0.05\textwidth]{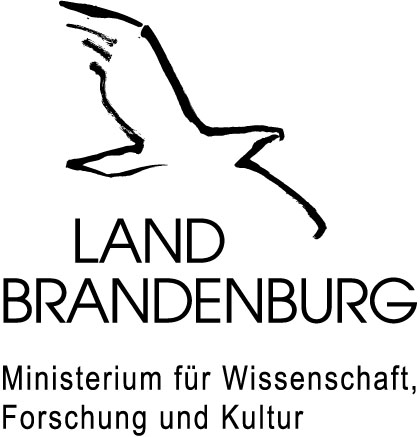}
\end{center}

\subsection*{Ethics declarations}
Not applicable.

\appendix
\onecolumngrid
\section{Details of the warm start procedures\label{app: detail-warm-start}}

For the WS-M approach, we have to maximize the expectation value in Eq.~\eqref{eq: f-local-op} by choosing appropriate variational parameters for the two-qubit gate acting on qubits $i$ and $j$. Utilizing the identities $ e^{-a \hat{P}} = \cosh(a) - \sinh(a) \hat{P}$ and $ e^{-ia \hat{P}} = \cos(a)-i\sin(a)\hat{P}$, for a real coefficient $a$ and $\hat{P}$ denoting one of the the Pauli operators, $\hat{P} \in \{I, \sigma^x, \sigma^y, \sigma^z\}^{\otimes 2}$, we can rewrite the overlap $f_{k}$ as follows
\begin{equation}\label{eq: f-detail}
    \begin{aligned}
        &f_{\tau, k}(\theta_{ij, 0}, \theta_{ij, 1}) \\
        &=  \bra{\psi_{k-1}} e^{-\tau \cdot J_{ij} \cdot \sigma_i^z \sigma_j^z} \cdot e^{-i (\theta_{ij,1} \cdot \sigma_i^z \sigma_j^y + \theta_{ij,0} \cdot \sigma_i^y \sigma_j^z)/2} \ket{\psi_{k-1}}\\
        &=\cos\left(\frac{\theta_{ij,1}}{2}\right)\cos\left(\frac{\theta_{ij,0}}{2}\right) \left(\cosh\left(\tilde{\tau}_{ij}\right) - \sinh\left({\tilde{\tau}_{ij}}\right)\cdot \langle \sigma_i^z \sigma_j^z \rangle \right)\\
        &-i\cos\left(\frac{\theta_{ij,1}}{2}\right)\sin\left(\frac{\theta_{ij,0}}{2}\right)\left(\cosh\left(\tilde{\tau}_{ij}\right)\langle \sigma_i^y \sigma_j^z \rangle +i \sinh\left({\tilde{\tau}_{ij}}\right)\cdot \langle \sigma_i^x \rangle \right)\\
        &-i\sin\left(\frac{\theta_{ij,1}}{2}\right)\cos\left(\frac{\theta_{ij,0}}{2}\right)\left(\cosh(\tilde{\tau}_{ij})\langle \sigma_i^z \sigma_j^y \rangle +i \sinh\left({\tilde{\tau}_{ij}}\right)\cdot \langle \sigma_j^x \rangle \right)\\
        &-\sin\left(\frac{\theta_{ij,1}}{2}\right)\sin\left(\frac{\theta_{ij,0}}{2}\right)\left(\cosh\left(\tilde{\tau}_{ij}\right)\langle \sigma_i^x \sigma_j^x \rangle + \sinh\left({\tilde{\tau}_{ij}}\right)\cdot \langle \sigma_i^y \sigma_j^y \rangle \right).
    \end{aligned}
\end{equation}
In the above expression, we have used the short-hand notations $\tilde{\tau}_{ij} = \tau J_{ij}$ and $\langle \hat{P} \rangle = \bra{\psi_{k-1}} \hat{P} \ket{\psi_{k-1}}$. Notice that $\langle \sigma_i^z \sigma_j^y \rangle = 0 =\langle \sigma_i^y \sigma_j^z \rangle$, because these two operators are imaginary and hermitian, resulting in a vanishing expectation in a real state. 

A similar calculation can be performed for the operators $\sigma_i^x\sigma_j^y, \sigma_i^y\sigma_j^x$ in the ansatz, the corresponding overlap will be given by
\begin{equation}\label{eq: xy-yx}
    \begin{aligned}
        &f^{\prime}_{\tau, k}(\theta_{ij, 0}, \theta_{ij, 1}) \\
        &=  \bra{\psi_{k-1}} e^{-\tau \cdot J_{ij} \cdot \sigma_i^z \sigma_j^z} \cdot e^{-i (\theta_{ij,1} \cdot \sigma_i^x \sigma_j^y + \theta_{ij,0} \cdot \sigma_i^y \sigma_j^x)/2} \ket{\psi_{k-1}}\\
        &=e^{-\tilde{\tau}_{ij}} \cdot \left(\cos\left(\frac{\theta_{ij,1}}{2}\right)\cos\left(\frac{\theta_{ij,0}}{2}\right) -  \sin\left(\frac{\theta_{ij,1}}{2}\right)\sin\left(\frac{\theta_{ij,0}}{2}\right) \cdot \langle \sigma_i^z \sigma_j^z \rangle \right).
    \end{aligned}
\end{equation}
This expression is maximized when $\theta_{ij,0} = 0= \theta_{ij,1}$, which explains why we did not include the operator $\sigma_i^x\sigma_j^y, \sigma_i^y\sigma_j^x$ in our ansatz.

Considering the expectation of the Pauli operators in Eq.~\eqref{eq: f-detail}, with operator $\hat{P} = \sigma_i^x, \sigma_j^x, \sigma_i^x \sigma_j^x,\sigma_i^y \sigma_j^y, \sigma_i^z \sigma_j^z$, the expectation $\langle \hat{P} \rangle$ can be expanded by using the definition of $\ket{\psi_{k-1}}$ from Eq.~\eqref{eq:ITE_two_qubit_terms}
\begin{equation}
    \begin{aligned}
        \langle \hat{P} \rangle &= \bra{\psi_{k-1}} \hat{P} \ket{\psi_{k-1}}\\
        &= \bra{\psi_{k-2}} e^{i (\theta_{pq,1} \cdot \sigma_p^z \sigma_q^y + \theta_{pq,0} \cdot \sigma_p^y \sigma_q^z)/2} \cdot \hat{P} \cdot e^{-i (\theta_{pq,1} \cdot \sigma_p^z \sigma_q^y + \theta_{pq,0} \cdot \sigma_p^y \sigma_q^z)/2}  \ket{\psi_{k-2}} \\
        &= \bra{\psi_{k-2}} \hat{P} \ket{\psi_{k-2}} + \bra{\psi_{k-2}} i\left[\sigma_p^z \sigma_q^y + \sigma_p^y \sigma_q^z, P\right] \ket{\psi_{k-2}} \cdot \mathcal{O}(\tau)  + \mathcal{O}(\tau^2), \\
        &=\  ^{N \otimes} \bra{+} \hat{P} \ket{+}^{\otimes N} + \left( \sum_{i\in V} \ ^{N \otimes} \bra{+} i\left[\sigma_i^y, P\right] \ket{+}^{\otimes N} + \sum_{pq \in E^{\prime}}\ ^{N \otimes} \bra{+} i\left[\sigma_p^z \sigma_q^y + \sigma_p^y \sigma_q^z, P\right] \ket{+}^{\otimes N} \right) \cdot \mathcal{O}(\tau) + \mathcal{O}(\tau^2)\\
        &= \  ^{N \otimes} \bra{+} \hat{P} \ket{+}^{\otimes N} + \mathcal{O}(\tau^2).
    \end{aligned}
\end{equation}
In the second line of the above equation, qubits indexed by $p, q$ refer to the gate applied at step $k-1$. The Baker-Campbell-Hausdorff formula, $e^{\hat{A}}\cdot \hat{P} \cdot e^{-\hat{A}} = \hat{P} + [\hat{A}, \hat{P}] + \frac{1}{2}[\hat{A}, [\hat{A}, \hat{P}]] \cdots $, was used for the expansion on the third line. Besides, it is assumed that the parameters $\theta_{pq, 0}$ and $\theta_{pq, 1}$ are of the same order as $\tau$. Finally, all gates existing in the circuit before step $k$ are expanded in a similar procedure. $E^{\prime}$ represents the pairs of two-qubit gates that were previously applied in the circuit, with $(p, q) \in E^{\prime}$ being different from $(i, j)$, such that the commutator $\left[\sigma_p^z \sigma_q^y + \sigma_p^y \sigma_q^z, P\right]$ will be zero or contain $\sigma^z$ or $\sigma^y$ operator, resulting in a zero expectation in state $\ket{+}^{\otimes N}$, similar to the terms $\left[\sigma_i^y, P\right]$. Therefore, the expectation $\langle \hat{P} \rangle$ can be estimated in the product state with the error $\mathcal{O}(\tau^2)$ .

\section{Numerical performance benchmarks of the SIA ansatz}\label{app: benchmark_ansatz}

\begin{figure}[htp!]
    \centering
    \includegraphics[width = 0.48\textwidth]{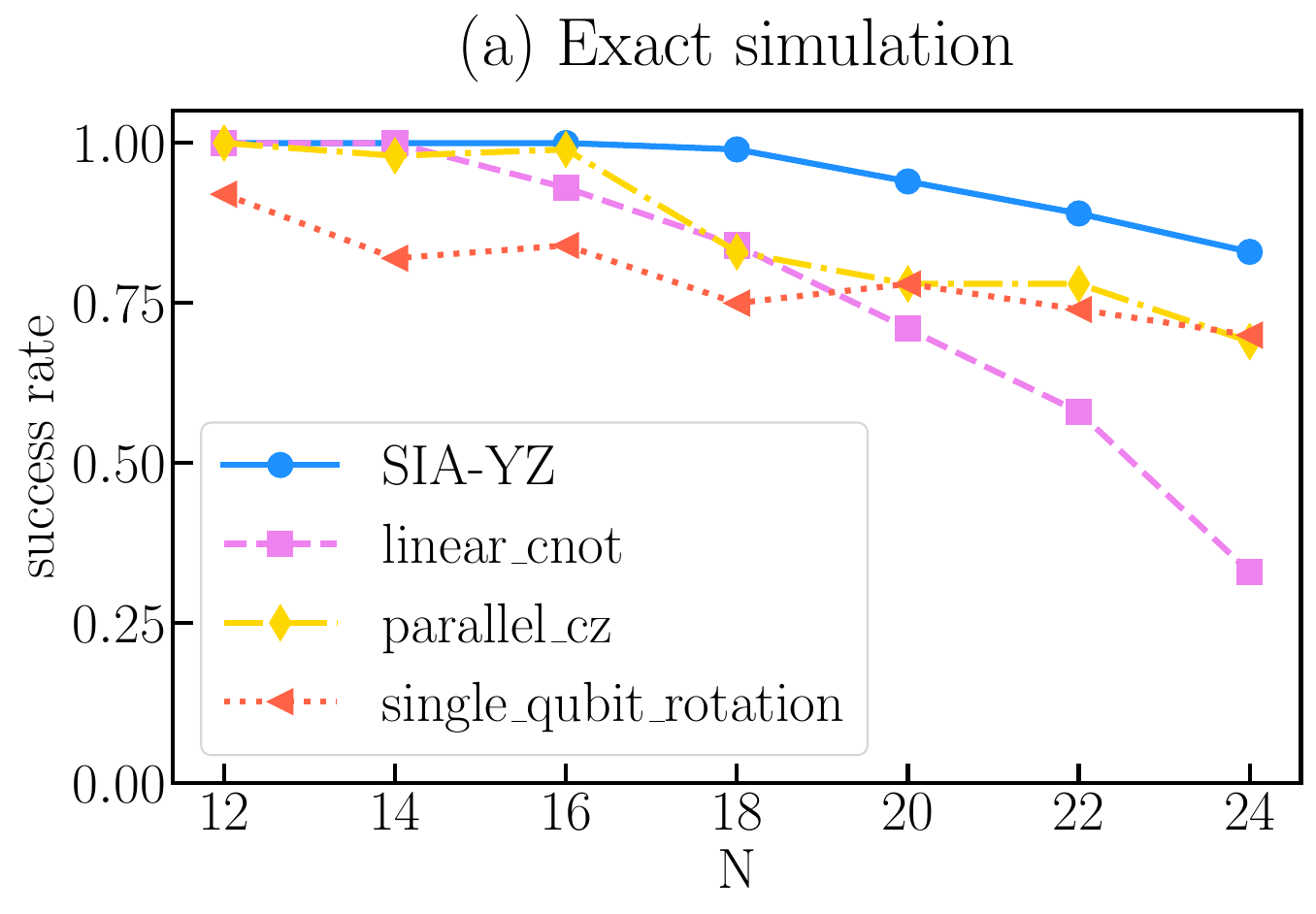}
    \includegraphics[width = 0.48\textwidth]{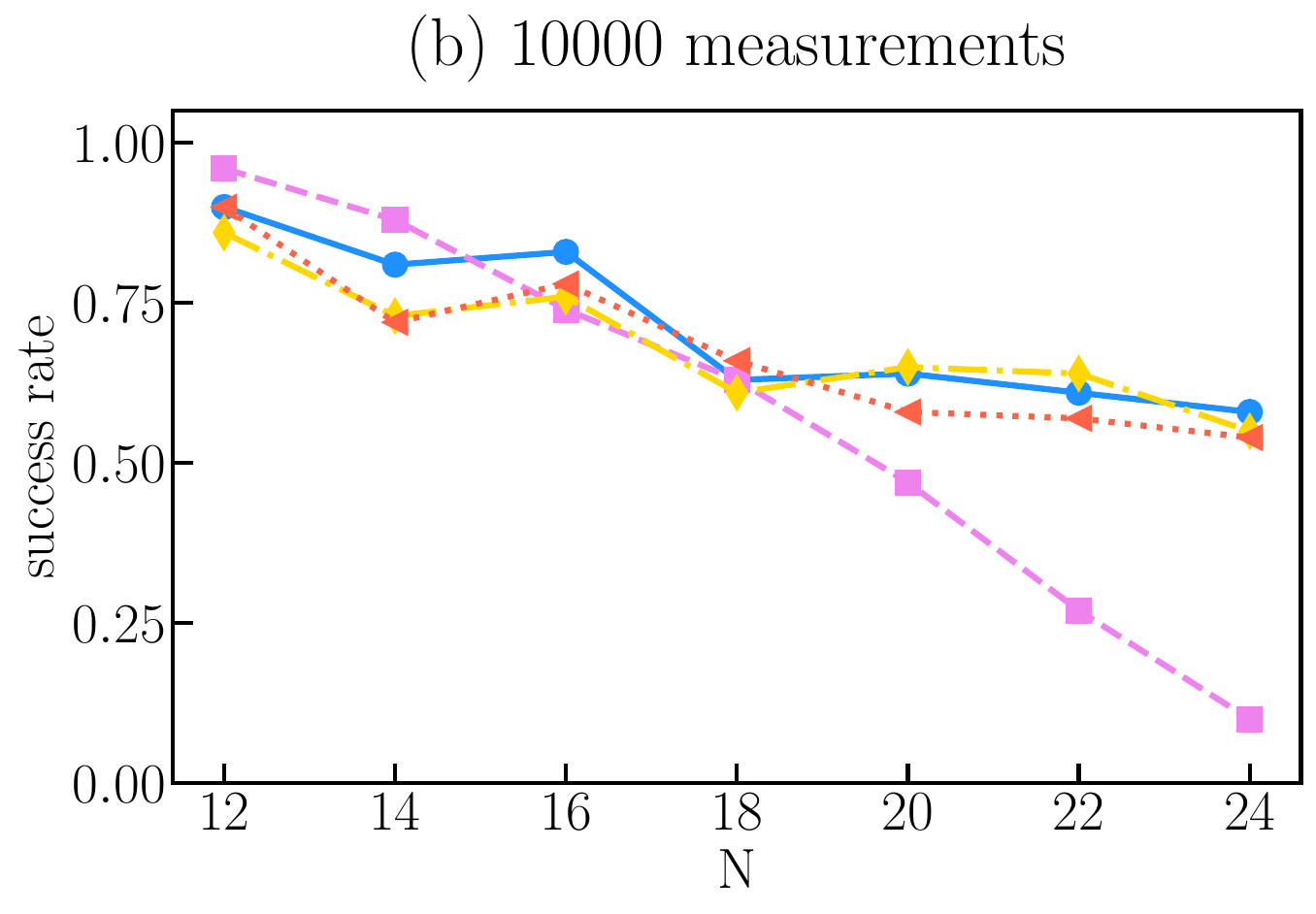}
    \caption{Success rate of CVaR-VQE with all parameters initialized as zero and $\alpha = 0.01$ for (a) exact state vector simulation and (b) a noise-free quantum device with 10000 measurements in every iteration. The green dot representst the SIA ansatz; the pink square is the HEA with linear CNOT as the entanglement layer; the orange diamond is the HEA with CZ gate as the entanglement layer; the red triangle is the single-qubit-rotation ansatz only produce product states.}
    \label{fig: success_rate}
\end{figure}

In this appendix, we benchmark the performance of the SIA ansatz compared with the HEA ansatz~\cite{Kandala_2017}, to explore if the SIA ansatz itself can lead to a better performance. Using the same instances as in the main text, Figure~\ref{fig: success_rate}, shows the success rate of the SIA introduced in the main text compared to various ansätze that haven been widely used in the literature.
In particular, we compare the SIA to HEAs with linear CNOT entanglement, parallel CZ entanglement and a single-qubit rotation ansatz that produces only product states (see Appendix~\ref{app: ansatzes}, Figs.~\ref{fig: hea_ansatz}(a) - \ref{fig: hea_ansatz}(c) for details).
For a fair comparison, all the ansätze were applied to the same initials state, the uniform superposition state $\ket{+}^{\otimes N}$, with the same number of parameters $N^2$ and all parameters set to zero initially, as detailed in Appendix.~\ref{app: ansatzes}. Note that we have not included the warm start for SIA yet.
Focusing on the results obtained from the exact simulation in Fig.~\ref{fig: success_rate}(a), corresponding to a perfect noise-free quantum computer with an infinite number of measurements, the SIA ansätz shows a slightly higher success rate than the other ansätze, in particular for large problem sizes.
The HEA ansatze with linear CNOT entangling layers performs only better than the single-qubit rotation ansatz without any entanglement for the problems with less than 18 qubits. From there on they are worse (HEA with linear CNOT entangling layers) or on par (HEA with parallel CZ entangling layers) with the single-qubit rotation ansatz.

Incorporating the fact that actual quantum devices only allow for a finite number of measurements changes the picture significantly, as Fig.~\ref{fig: success_rate}(b) reveals. Despite using a relatively large number of 10000 measurements per iteration, the statistical noise leads to a significant drop in performance of the SIA ansätze. In particular, they lose their advantage over the single-qubit rotation ansatz, and except for the HEA with linear CNOT entangling layers, all variational ansätze show very similar performance for the entire range of problem sizes we study. Our results are in agreement with the conclusion of Ref.~\cite{role_entangle}, which found that entanglement gates adapted to the problem structure are beneficial for optimization problems while the advantage will decrease in the case of a finite number of measurements. The poor performance of HEA and SIA at large problem sizes with a finite number of measurements might be the result of statistical errors as well as barren plateaus, which is discussed in Sec.~\ref{sec: discussion}. Based on this observation, we conclude that the SIA ansatz alone can not provide practical performance improvement, highlighting the importance of finding a good initial parameter to enhance VQE.

\section{Additional tests for HEA with warm start initialization}
\label{app:hea_ws_appendix}

For completeness, we also examined whether a warm start procedure can improve the performance of the HEA. Since HEA does not incorporate any problem structure, identifying an efficient warm start is challenging. Nevertheless, following the approach used in Ref.~\cite{Egger2021warmstartingquantum}, we constructed a warm start based on the continuous relaxation of the QUBO problem. For each instance, we solve 
\begin{equation}
    \min_{x\in[0,1]^N} x^T Q x,
\end{equation}
where the binary variables $x_i\in\{0,1\}$ are relaxed to $x_i\in[0,1]$. The optimal relaxed solution, denoted by $c_i^\ast$, can be efficiently obtained via exact diagonalization of the $N\times N$ matrix $Q$. We then initialize each qubit in the state $\ket{\phi_i} = R_y(\theta_i)\ket{0}$ with
\[
    \theta_i = 2\arcsin\!\left(\sqrt{c_i^\ast}\right),
\]
such that the probability of measuring $\ket{1}$ equals $c_i^\ast$. 
(As all ansatze begin with a Hadamard gate in our implementation, this is equivalent to applying $R_y(\theta_i - \pi/2)$ after a Hadamard on $\ket{0}$.)
To avoid extreme cases $c_i^\ast = 0$ or $1$, we follow Ref.~\cite{Egger2021warmstartingquantum} and introduce a regularization parameter $\varepsilon = 0.25$, resulting in the initialization rule
\begin{equation}
\theta_i =
\begin{cases}
2 \arcsin\!\left(\sqrt{c_i^\ast}\right), & c_i^\ast \in [\varepsilon, 1-\varepsilon],\\[6pt]
2 \arcsin\!\left(\sqrt{\varepsilon}\right), & c_i^\ast \le \varepsilon,\\[6pt]
2 \arcsin\!\left(\sqrt{1-\varepsilon}\right), & c_i^\ast \ge 1-\varepsilon.
\end{cases}
\end{equation}

\begin{figure}[t]
    \centering
    \includegraphics[width=0.96\textwidth]{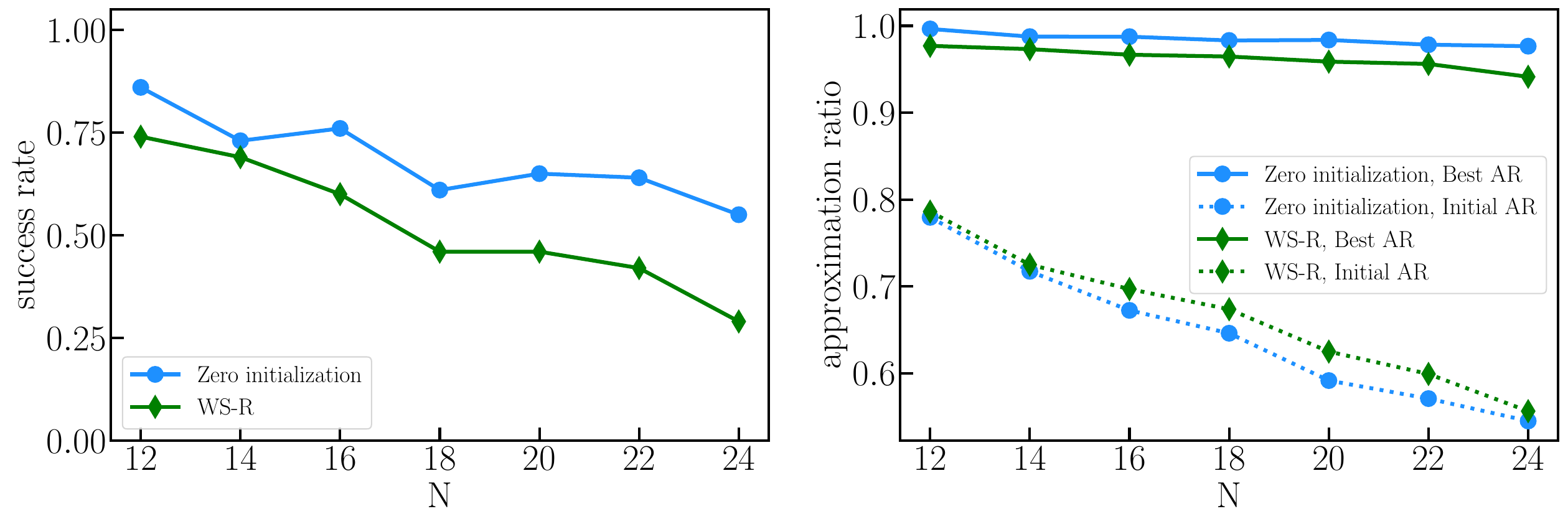}
    \caption{Performance of CVaR-VQE using the HEA with zero initialization and WS-R. 
    The entangling pattern is the parallel CZ entangler shown in Fig.~\ref{fig: hea_ansatz}(b). 
    We use $\alpha = 0.01$ and $10000$ shots. 
    (a) Success rate: blue dots correspond to zero initialization and green diamonds to WS-R. 
    (b) Approximation ratio: dashed lines show the ratio at initialization, while solid lines show the best ratio obtained during CVaR-VQE.}
    \label{fig:HEA-WS}
\end{figure}

We use these parameters to initialize the first-layer parameters of the HEA with parallel CZ entanglement (Fig.~\ref{fig: hea_ansatz}(b)), while all remaining parameters are set to zero. We refer to this method as \emph{warm start by relaxation} (WS-R). Using WS-R as the initialization for the HEA, we performed CVaR-VQE simulations with $\alpha = 0.01$ and $10{,}000$ shots. As shown in Fig.~\ref{fig:HEA-WS}(a), the success rate does not improve compared to zero initialization. To confirm that WS-R indeed provides a better starting state, Fig.~\ref{fig:HEA-WS}(b) compares the approximation ratio at initialization and the best approximation ratio obtained during optimization. Although WS-R yields a higher initial approximation ratio, it does not improve the subsequent VQE convergence. This may be because the HEA does not reflect the problem structure and thus cannot effectively exploit the warm start state. A similar conclusion was drawn in Ref.~\cite{Egger2021warmstartingquantum}: as shown in Fig.~4(b) of that work, the warm start is beneficial only when the mixer of QAOA is adapted to the warm start state; otherwise, using the warm start state with the standard mixer does not improve performance.

 To confirm that this behavior is not specific to the parallel CZ entanglement, we also tested HEA with linear CNOT and single-qubit rotation. In both cases, we observed the same qualitative behavior: WS-R improves the initial approximation ratio but does not enhance the overall optimization performance.  Since these additional results do not provide further qualitative insights beyond Fig.~\ref{fig:HEA-WS}, we omit the corresponding plots from the main text. The findings support our conclusion that a warm start strategy is effective only when the ansatz is aligned with the problem or the initialization state, as is the case for the SIA, but not for problem-agnostic ansatze such as the HEA.

\section{Circuits for different ansätze\label{app: ansatzes}}

\begin{figure}[htp!]
    \centering
    \includegraphics[width = 0.48\textwidth]{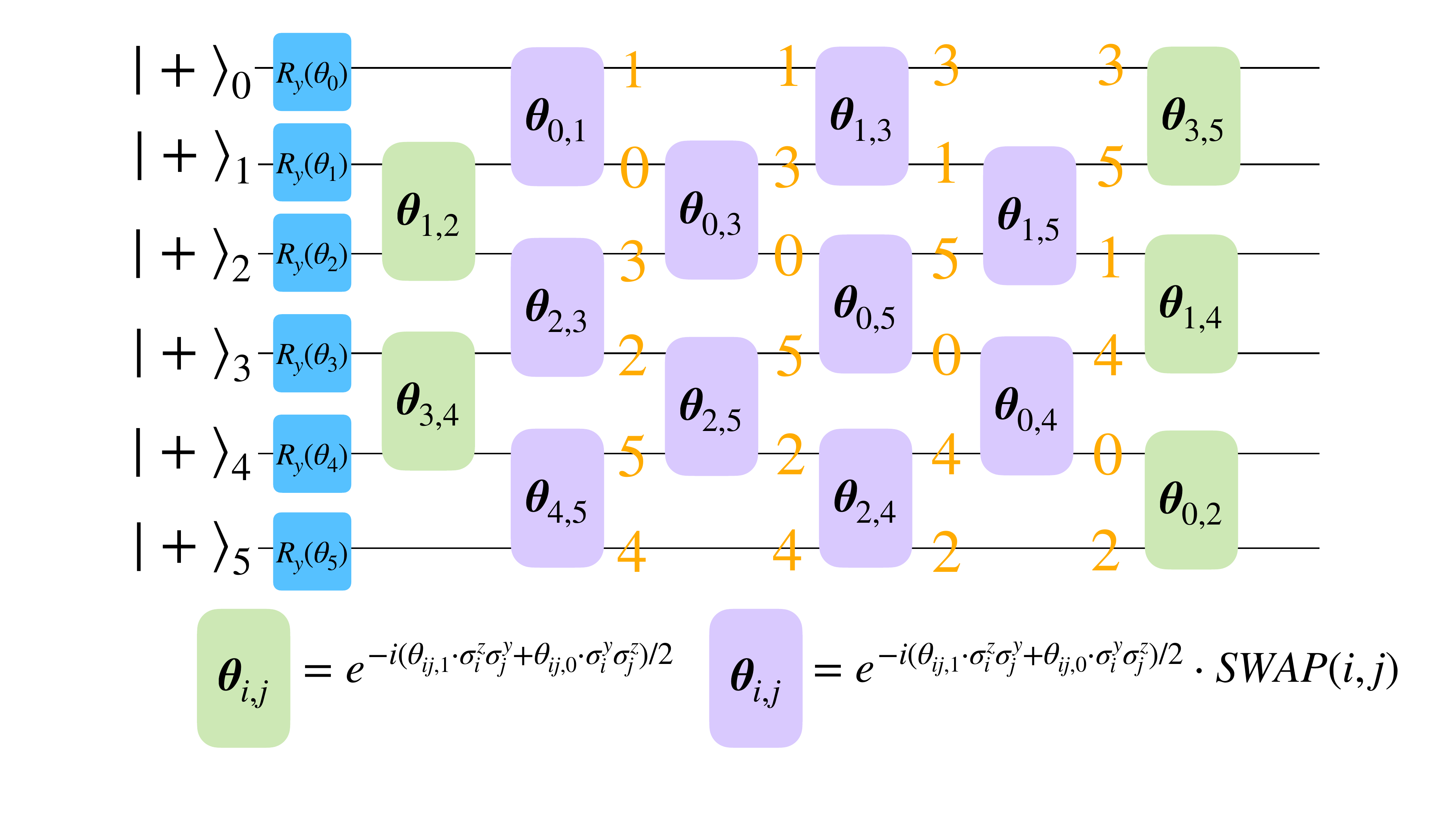}
    \caption{SIA circuit for 6 qubits. The yellow number indicates the qubit index after the SWAP operation.}
    \label{fig: SIA_circuit}
\end{figure}

The circuit for the SIA-YZ ansatz includes two-qubit gates for all possible qubit pairs. In general, current hardware platforms only offer limited qubit connectivity, and we need additional SWAP gates to realize all-to-all connectivity. Following the strategy in Ref.~\cite{scaling_qaoa_swap}, we take 6 qubits as an example and show the resulting circuit assuming linear qubit connectivity in Fig.~\ref{fig: SIA_circuit}. Full connectivity is reached after $N-2$ SWAP layers, and the SWAP gate is compressed with the operator $U_{ij}(\boldsymbol{\theta}_{ij})^{YZ}$ in Eq.~\eqref{eq: SIA-YZ}, thus they can be implemented with 3 CNOT gates (purple box in Fig.~\ref{fig: SIA_circuit}). Additionally, the first and final layers are only gates $U_{ij}(\boldsymbol{\theta}_{ij})^{YZ} \in SO(4)$ that can be decomposed by using 2 CNOT gates~\cite{Vatan_2004} (green box in Fig.~\ref{fig: SIA_circuit}). Consequently, the overall structure requires $3N-2$ CNOT layers and $\mathcal{O}(1.5 N^2)$ CNOT gates. Alternatively, if the quantum device has all-to-all connectivity, no additional SWAP gates are necessary and the circuit only consists of the gates $U_{ij}(\boldsymbol{\theta}_{ij})^{YZ}$ and single-qubit rotation gates. In this case the circuit depth is reduced to $2N$ CNOT layers and $\mathcal{O}(N^2)$ CNOT gates.

\begin{figure}[htp!]
    \centering 
    \subfigure[\ Hardware-efficient ansatz with linear CNOT entanglement.]{
    \label{Fig.linear_cnot}
    \includegraphics[width = 0.35\textwidth]{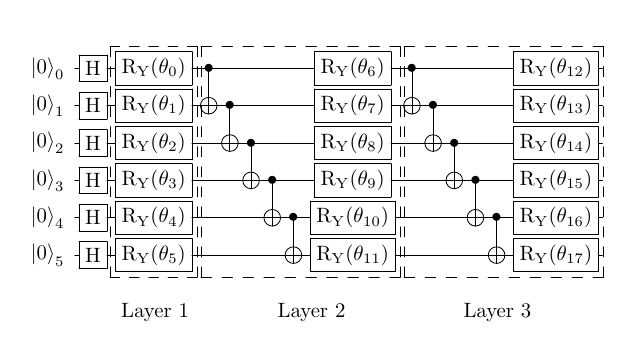}}
    \subfigure[\ Hardware-efficient with parallel CZ entanglement.]{
    \label{Fig.parallel_cz}
    \includegraphics[width = 0.3\textwidth]{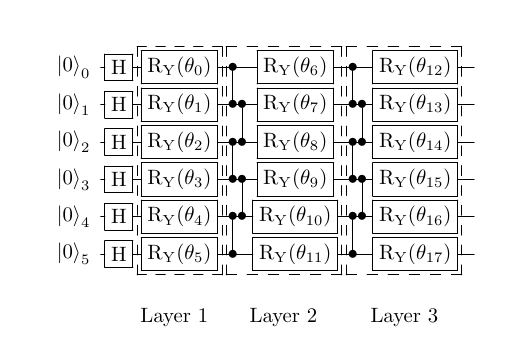}}
    \subfigure[\ Single-qubit-rotation ansatz]{
    \label{Fig.single_qubit_rotation}
    \includegraphics[width = 0.3\textwidth]{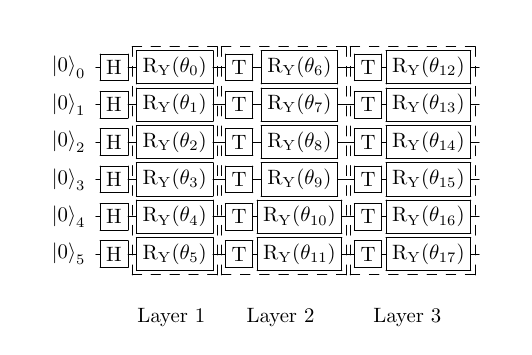}}
    \caption{Different ansätze used for comparison with the SIA. Panel (a) shows the hardware-efficient ansatz producing real amplitudes with linear CNOT entangling layers, panel (b) the same ansatz but with non-overlapping, parallel CZ entangling layers and panel (c) an ansatz consisting of single-qubit gates only, thus not being able to generate any entanglement.}
    \label{fig: hea_ansatz}
\end{figure}

In addition to the SIA, we consider three alternative ansätze often used in the literature for comparison, which are illustrated in Fig.~\ref{fig: hea_ansatz}. For a fair comparison, all ansätze have the same number of parameters, utilizing $N^2$ parameters equivalent to those of a single-layer SIA, thereby necessitating $N$ layers for the ansatz depicted in Fig.~\ref{fig: hea_ansatz}. Moreover, Table~\ref{table: cnots_ansazes} shows the number of two-qubit gates required for each ansatz, maintaining the same parameter count as the 1-layer SIA.

\begin{table}[!htbp]
    \centering
    \begin{tabular}{|c|c|c|c|c|}
    \hline
    \diagbox{qubit connectivity}{ansatz} & SIA & HEA with linear CNOT & HEA with parallel CZ & Single-qubit rotations\\ 
    \hline
    linear           &$1.5N^2 - 2.5N + 1$   & $(N-1)^2$ & $(N-1)^2$ & 0\\
    \hline
    all-to-all       &$N(N-1)$   & $(N-1)^2$ & $(N-1)^2$ & 0\\
    \hline
    \end{tabular}
    \caption{Number of CNOT gates for different ansätze with $N^2$ parameters, with respect to linear or all-to-all qubit connectivity of the hardware platform.}
    \label{table: cnots_ansazes}
\end{table}

\section{Performance of the warm start approach for more than a single layer\label{app: layers}}
\begin{figure}[H]
    \centering
    \includegraphics[width = 0.5\textwidth]{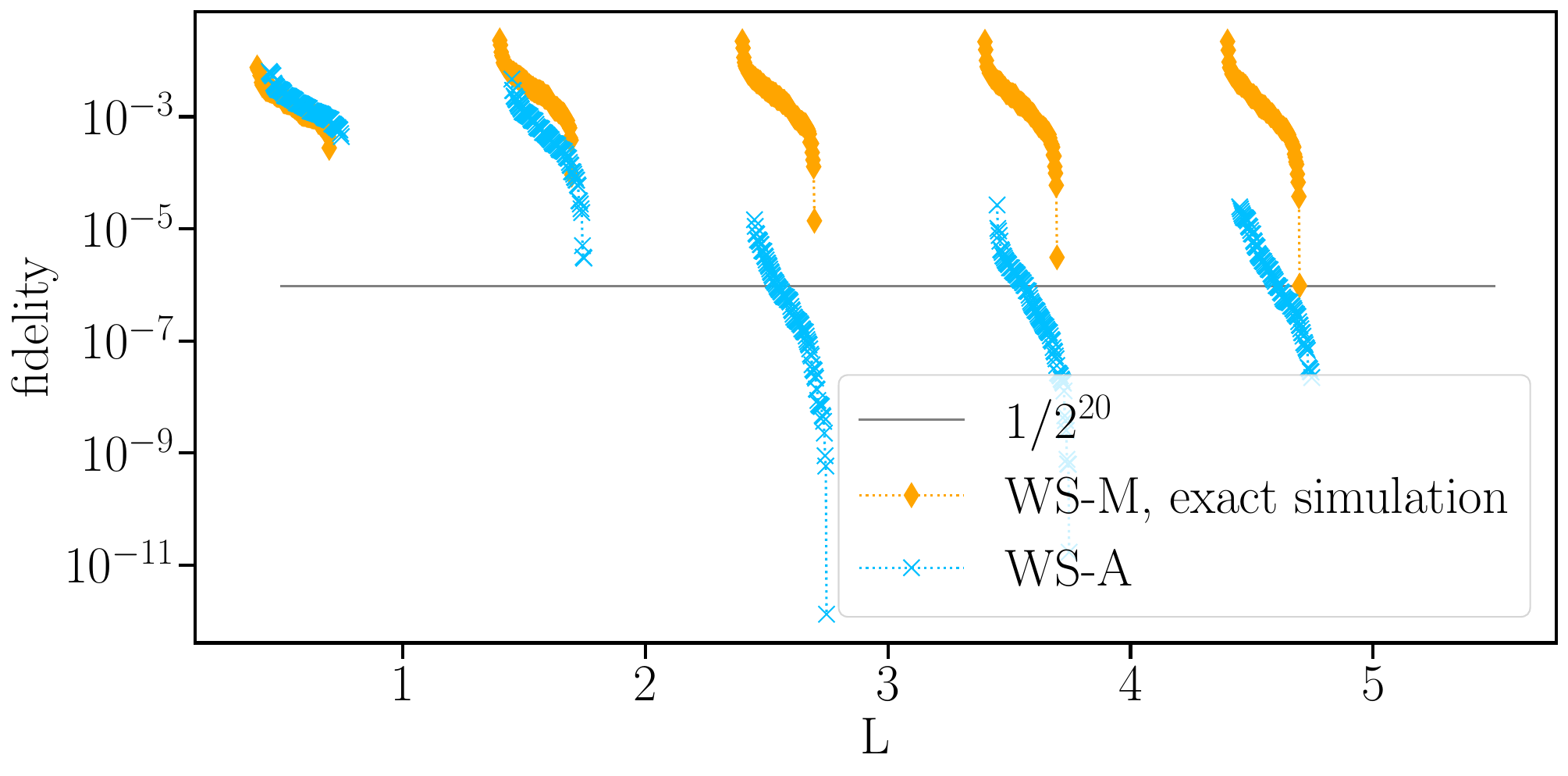}
    \caption{Fidelity of WS-M and WS-A methods using a state vector simulation for 20 qubits for different numbers of layers, $\tau = 0.2$ is used for all data points.}
    \label{fig: warm_start_fidelity_layers}
\end{figure}

This appendix shows results for the performance of the warm start across different number of layers, $L = 1, 2, 3, 4, 5$. Figure~\ref{fig: warm_start_fidelity_layers} shows the data for 20 qubits as an example and compares the fidelity between the WS-M and WS-A. While both approaches yield comparable results for a single layer, the fidelity achieved through the WS-A decreases with an increasing number of layers. Conversely, the WS-M method maintains a relatively consistent performance for all values of $L$ we consider. Notably, the fidelity improves slightly when going from a single layer to $L=2$, but there is no significant improvement when further adding additional layers. This observation suggests the necessity to incorporate more non-local operators if one wants to enhance the fidelity of the initial state with the exact solution. 

\twocolumngrid
\bibliography{references}
\end{document}